\documentclass[authoryear, preprint,12pt]{elsarticle}
\usepackage{amsmath}

\usepackage{amssymb}

\newtheorem{prop}{Proposition}
\usepackage[colorlinks=true,citecolor=blue, linkcolor=blue]{hyperref}
\usepackage[normalem]{ulem}
\usepackage{subcaption}

\usepackage{xcolor}
\usepackage{float}
\usepackage{appendix}
\journal{arXiv}
\begin{document}

\begin{frontmatter}
\title{Quantifying optimal resource allocation strategies for controlling epidemics}

\author[inst1]{Biplab Maity}
\author[inst1,inst3]{ Swarnendu Banerjee\corref{corauthor}\fnref{firstfoot}}
\author[inst1,inst2]{Abhishek Senapati\corref{corauthor}\fnref{firstfoot} }
\author[inst1]{Joydev Chattopadhyay}
\cortext[corauthor]{Corresponding author\\ Email: abhishekiz4u04@gmail.com;
swarnendubanerjee92@gmail.com}

\address[inst1]{Agricultural and Ecological Research Unit, Indian Statistical Institute, 203, B. T. Road, Kolkata 700108, India}
\address[inst2]{Center for Advanced Systems Understanding (CASUS), Goerlitz, Germany}
\address[inst3]{Copernicus Institute of Sustainable Development, Utrecht University, PO Box 80115, 3508 TC, Utrecht, The Netherlands}
\fntext[firstfoot]{Equal contribution}



\begin{abstract}
Frequent emergence of communicable diseases has been a major concern worldwide. Lack of sufficient resources to mitigate the disease-burden makes the situation even more challenging for lower-income countries. Hence, strategy development towards disease eradication and optimal management of the social and economic burden has garnered a lot of attention in recent years. In this context, we quantify the optimal fraction of resources that can be allocated to two major intervention measures, namely reduction of disease transmission and improvement of healthcare infrastructure. Our results demonstrate that the effectiveness of each of the interventions has a significant impact on the optimal resource allocation in both long-term disease dynamics and outbreak scenarios. Often allocating resources to both strategies is optimal. For long-term dynamics, a non-monotonic behavior of optimal resource allocation with intervention effectiveness is observed which is different from the more intuitive strategy recommended in case of outbreaks. Further, our result indicates that the relationship between investment into interventions and the corresponding outcomes has a decisive role in determining optimal strategies. Intervention programs with decreasing returns promote the necessity for resource sharing. Our study provides a fundamental insight into determining the best response strategy in case of controlling epidemics under resource-constrained situations.
\end{abstract}

\end{frontmatter}



    



\section{Introduction}

Infectious diseases like Influenza, Severe Acute Respiratory Syndrome (SARS), Ebola, and most recently COVID-19 spread throughout a population within a short time making it a cause of utmost concern for human society \citep{fraser2009pandemic, colizza2007predictability, hufnagel2004forecast,legrand2007understanding, zhang2020changes,mahase2020china,gubler2002epidemic}. Lack of appropriate health policies, protective medical equipment such as disposable gloves, surgical masks and so on, poorly developed infrastructure, and shortage of medical personnel make it even more challenging for low-income countries to mitigate the health crisis during an ongoing outbreak \citep{oshitani2008major,hopman2020managing}. This can be well understood, for instance, by considering the severe impact of Ebola on West African countries like Nigeria, Liberia, and Sierra Leone where there was an estimated shortage of 7.2 million doctors and health workers \citep{gostin2014ebola, gostin2015retrospective}. Furthermore, estimates also indicate that nearly $79\%$ of total available hospital beds would be required for patients due to Influenza in the lower income countries with an incidence rate of $35\%$. This estimate becomes more than $100\%$ in countries like Bangladesh and Nepal even at the incidence rate of $15\%$ \citep{oshitani2008major}. Therefore, the allocation of resources in an optimal way becomes a necessary component of public health response toward controlling the spread of infectious diseases. 

A simplistic way to look at control of infectious disease outbreaks is either by prevention of transmission through non-pharmaceutical interventions (NPI) such as social distancing, personal hygiene, contact tracing, and early detection or improvement of recovery rate by providing better medical treatment. World Health Organization \citep{world2005checklist, world2020coronavirus} and governmental health agencies in many countries publish disease prevention protocols to fight against disease outbreaks \citep{benson2009guidelines,siegel20072007, adhikari2020epidemiology}. There are also campaigns, often using various social networking websites and educational programs to increase awareness among people about a particular disease. In absence of a proper effective vaccine or vaccination strategy, the implementation of various preventive measures helps to diminish the contact among the infected and susceptible individuals thereby reducing the transmission rate. Although NPIs are likely to be effective at early stages but often it is not possible to keep the disease transmission under control, especially, when there is a sudden flare-up of cases. Additionally, there are also difficulties in implementing control measures like lockdown or travel restrictions due to huge economic and societal costs, mainly in developing countries~\citep{angulo2021simple, ngonghala2020mathematical}. In such a case, it is important to ramp up hospital infrastructure and consequently prioritize the quality of treatment which would influence the recovery of infected individuals thus reducing the potential impact of an epidemic \citep{farmer2001major, mosadeghrad2014factors, emanuel2020fair}. Such measures would include increasing the number of hospital beds \citep{sacchetto2020covid}, setting up temporary medical units, training program of healthcare personnel \citep{oshitani2008major}, and procuring more medical equipment and medicines. 

Designing efficient and feasible control strategies that maintain a balance between the implementation of pharmaceutical as well as non-pharmaceutical interventions in a resource-constrained situation is essential to reduce the potential impact of disease burden. However, in both cases, effectiveness of the interventions plays a key role in determining the success of the implemented strategies. One intervention measure is said to be more effective than the other, if a better recovery rate or a reduced transmission rate is achieved at the expense of the same resource. For instance, in order to mitigate Ebola and SARS, symptom monitoring program becomes a more effective measure than quarantine, whereas the benefit of quarantine over symptom monitoring has been observed for diseases like Influenza and smallpox \citep{peak2017comparing}. Also, face covering appears to be a more effective way to prevent COVID-19 than that of quarantine and maintaining social distancing measures \citep{zhang2020identifying}. On the other hand, treatment protocol using drugs like Remdesivir was shown to be more effective than that of Hydroxychloroquine during the early stages of the COVID-19 pandemic \citep{beigel2020remdesivir,recovery2020effect}. Furthermore, the relationship between invested resources and the corresponding outcome from an intervention program, also known as the production function, plays a decisive role in determining optimal strategies \citep{brandeau2005improved}. Since the outcome of an intervention depends on its effectiveness, the later influences the steepness of a particular type of production function.
 Hence, it becomes important to design resource allocation policies by taking into account the effectiveness of available intervention programs and {their corresponding production functions to} evaluate their performance in disease mitigation.

 To analyse a range of possible scenarios under specific epidemic situations, mathematical models have been used as an important and reliable tool to guide policy making and implementing different intervention programs. In this regard, several attempts have been made that investigated different aspects of optimal resource allocation strategies \citep {medlock2009optimizing, alistar2014hiv,brandeau2009optimal, hansen2011optimal, worby2020face, brandeau2003resource, senapati2019impact, calabrese2022optimal, ghosh2021optimal, bolzoni2019optimal}.  
Many of these studies have focused on resource allocation to only prevention programs related to vaccination \citep{medlock2009optimizing}, or transmission reductions like face mask distribution \citep{worby2020face} and testing \citep{calabrese2022optimal,ghosh2021optimal}. Few studies have also considered resource allocation between two different strategies such as vaccination and isolation but did not take into account any explicit trade-off between them \citep{hansen2011optimal,bolzoni2019optimal}. Thus these studies lack in providing a theoretical framework to study optimal resource allocation between two main classes of intervention strategies - transmission reduction and improving healthcare infrastructure where there can be trade-offs between the two. While a recent study examined the allocation of resources between prevention and treatment, the role of effectiveness of the former was not explored \citep{alistar2014hiv}. Moreover, the study was based on HIV transmission and hence did not consider a recovered class which might be relevant for other diseases. For the same reason, the optimal strategy was determined considering only specific types of production functions.

In this study, we fill this gap and attempt to provide a general understanding of how limited resources may be allocated between transmission-reducing and recovery-improvement programs taking into account the effectiveness of both the types of interventions. Using a simple SIR model, we quantify the optimal fraction of resources required to minimize the impact of disease burden under both long-term disease dynamics and outbreaks. We study the role of different types of production functions and explored how the optimal strategy varies with the effectiveness of the intervention measures. Furthermore, since there is a delay in reporting new infections and proper knowledge about the disease is not available at early stages, we also quantify the optimal fraction of resources in the presence of delay in implementing interventions.














\section{Methods}
\subsection{Model}\label{sec:model_formulation}
We consider the well-known compartmental Susceptible-Infected-Recovered (SIR) model~\citep{anderson1992infectious, breda2012formulation} to represent the disease dynamics. At time $t$, the fractions of susceptible, infected, and recovered population are denoted by $S(t)$, $I(t)$, and $R(t)$ respectively which satisfies the constraint $S(t) + I(t) + R(t)=1.$ The rate of change of the fraction of people in each compartment can be described as follows:
\begin{eqnarray}
\begin{array}{lll}

\displaystyle \frac{dS}{dt} &=& \displaystyle \mu- \beta_0 SI- \mu S,\\\\

\displaystyle \frac{dI}{dt} &=& \displaystyle \beta_0 SI- \gamma_0 I - \mu I,\\\\

\displaystyle \frac{dR}{dt} &=& \displaystyle  \gamma_0 I-\mu R.\\\\

\end{array}
\label{basic_model}
\end{eqnarray}

Here, $\beta_0$ is the disease transmission rate, $\mu$ is the birth and death rate, and $\gamma_0$ is the recovery rate. The basic reproduction number, $\mathcal{R}_0$, defined as the average number of secondary cases appearing from an average primary case in an entirely susceptible population, can be obtained as $\mathcal{R}_0$=$\displaystyle \frac{\beta_0}{\gamma_0+\mu}.$
    
\begin{figure}[H]
    	\begin{center}
    		 \includegraphics[width=0.8\textwidth]{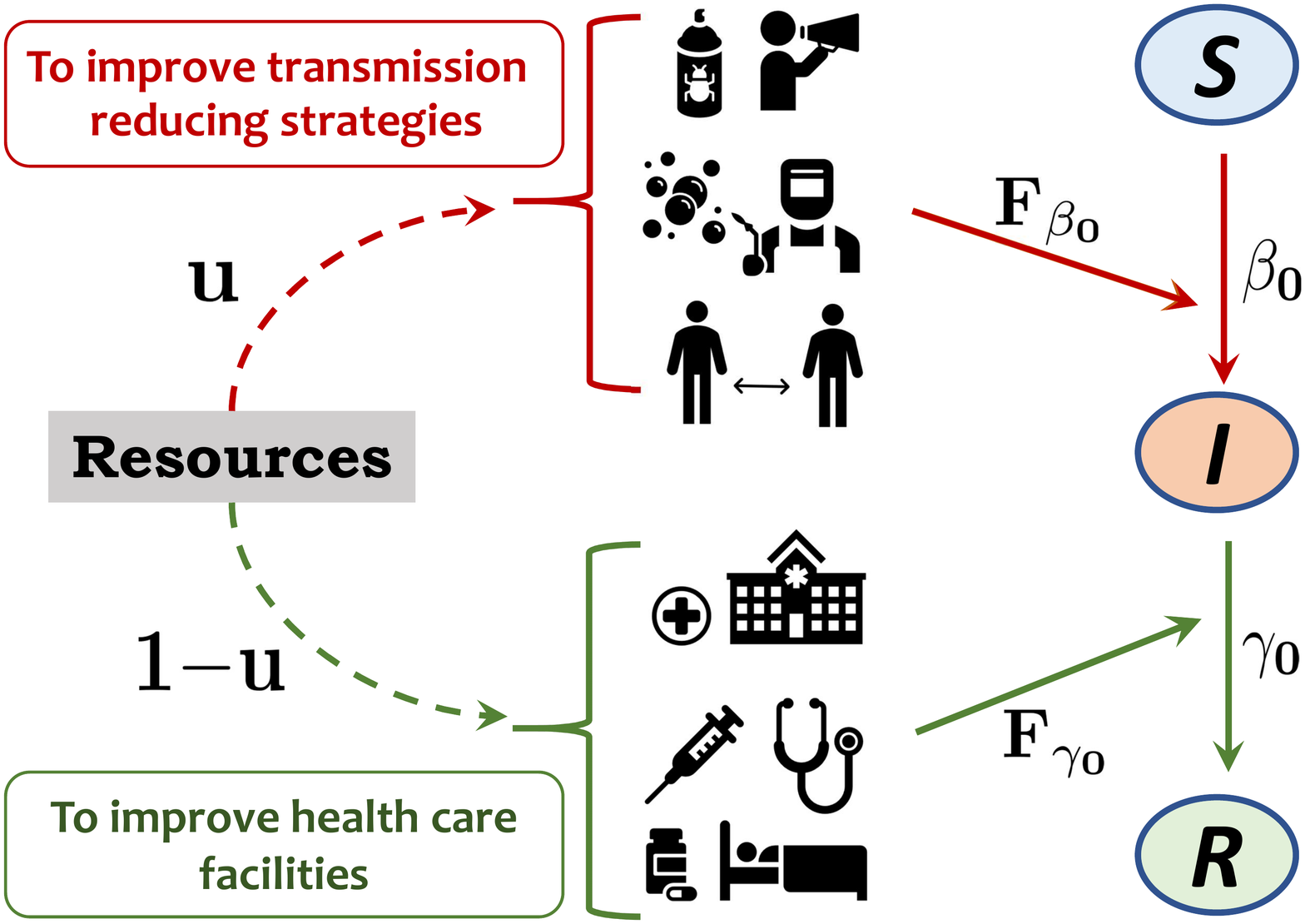}
    	\end{center}
    	\caption{\rm Optimal resource allocation. The total resources available for disease control are allocated to two broad classes of interventions namely transmission-reducing interventions like awareness programs, personal hygiene, social distancing, etc., and the intervention programs that would improve health care facilities which in turn shortens the infectious period and speed up the process of recovery. Here, $u$ is the fraction of the total available resources allocated to the transmission-reducing interventions and the remaining $(1-u)$ fraction is allocated to improve healthcare facilities. {This process of resources allocation is incorporated in the classical SIR model by introducing the production functions, $F_{\beta_{0}}$ and $F_{\gamma_{0}}$, that will reflect the effect of resource allocation in transmission rate ($\beta_0$) and recovery rate ($\gamma_0$) respectively.}
    	}
    	\label{schematic}
    \end{figure}

We introduce control strategies utilizing available resources among competing prevention and treatment programs. Here, one can think of the available resources as the total financial budget for government investment in healthcare. A natural consequence of the allocation of resources is an alteration in disease transmission rates and recovery rates that affect epidemic outcomes.  Here, a fraction $u$ of total available resources is allocated for implementing transmission-reducing strategies, and the remaining fraction, i.e., ($1-u$) is devoted to improving healthcare infrastructure (see the schematic in Fig.~\ref{schematic}). The fact that the total available resources to reduce disease burden are fixed, signifies a trade-off between allocations to the two types of competing intervention programs.

The relationship between the invested resources and outcomes of intervention programs, described by the production function, determines the true impact of public health interventions. We denote $F_{\beta_{0}}(u)$ and $F_{\gamma_{0}}(1-u)$ as the production functions associated with transmission-reducing and recovery-improvement programs, respectively. To reflect the effect of resource implementation, we replace $\beta_{0}$ and $\gamma_{0}$ in model \eqref{basic_model} by $F_{\beta_{0}}(u)$ and $F_{\gamma_{0}}(1-u)$ respectively (see the schematic in Fig.~\ref{schematic}). So the model with interventions can be written as 

\begin{eqnarray}
\begin{array}{lll}
\displaystyle \frac{dS}{dt} &=& \displaystyle \mu- F_{\beta_{0}}(u) SI- \mu S,\\\\
\displaystyle \frac{dI}{dt} &=& \displaystyle F_{\beta_{0}}(u)SI - F_{\gamma_{0}}(1-u)I - \mu I,\\\\
\displaystyle \frac{dR}{dt} &=& \displaystyle  F_{\gamma_{0}}(1-u)I-\mu R.\\
\end{array}
\label{model_with_intervention}
\end{eqnarray}

\begin{figure}[H]
    	\begin{center}
    		 \includegraphics[width=0.8\textwidth]{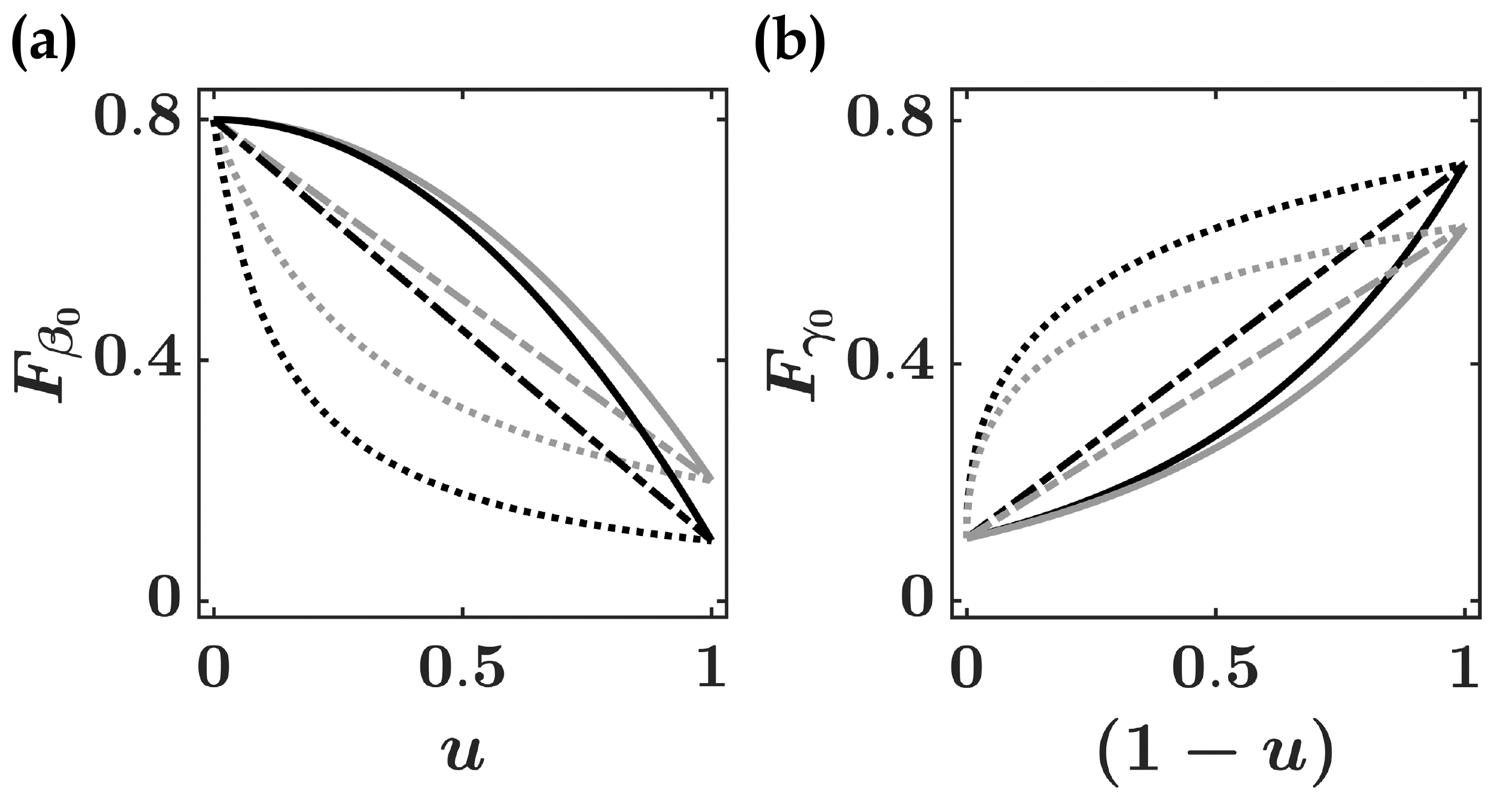}
    	\end{center}
    	\caption{\rm {Different types of production functions: (a) $F_{\beta_{0}}(u)$ and (b) $F_{\gamma_{0}}(1-u)$ corresponding to transmission-reducing and recovery-improvement interventions respectively. Here solid, dashed, and dotted lines represent increasing, constant and decreasing returns to scale respectively. In each of (a) and (b), the production functions with lower and higher effectiveness parameters are presented in grey and black colour respectively. For specific forms of $F_{\beta_{0}}$ and $F_{\gamma_{0}}$, refer to~\ref{Appendix-B}, Fig.~\ref{outbreak_9_figure} caption.}}
    	\label{production_function_figure}
    \end{figure}

Now, production functions, $F_{\beta_{0}}$ and $F_{\gamma_{0}}$, can be broadly categorized into three types as described below. For each incremental investment in the intervention programs, the associated production function, $F_{\beta_{0}}$ $ (F_{\gamma_{0}})$ gives decreasing, constant, or increasing returns to scale if the corresponding reduction (improvement) in the transmission (recovery) rate is increasingly smaller, remains constant or increasingly larger respectively (see Fig.~\ref{production_function_figure}) \citep{brandeau2003resource,brandeau2009optimal}. 
For the sake of comparison, we assume that three types of returns considered for $\displaystyle F_{\beta_0}$ ($\displaystyle F_{\gamma_0}$) are equivalent to each other at the two extreme cases, i.e., when $u = 0$ or $u = 1$ (see Fig.~\ref{production_function_figure}). Further, it is important to note that there is always a limited realizable benefit from an intervention program regardless of investment \citep{brandeau2005improved,brandeau2003resource}. This implies that there always will be a lower limit to the reducible transmission rate and an upper limit to the achievable recovery rate. However, these limits are dependent on the effectiveness of the intervention strategies. For the same resources, a more effective intervention measure implies more reduction (increase) in transmission (recovery) rate (see Fig.~\ref{production_function_figure}).

We consider {the production functions}, $\displaystyle F_{\beta_{0}}(u)=\displaystyle \frac{\beta_{0}}{1+\beta_{e}u}$, and $F_{\gamma_{0}}(1-u)=\gamma_{0}(1+\gamma_{e}(1-u))$, where the parameter $\beta_{e}$ indicates the effectiveness of transmission-reducing interventions, and the parameter $\gamma_{e}$ denotes the effectiveness of recovery-improving interventions. Here, $F_{\beta_0}$ and $F_{\gamma_0}$ have decreasing and constant returns to scale, respectively. The motivation for such consideration comes from several theoretical as well as empirical studies on resource allocation for HIV prevention programs like needle exchange \citep{kaplan1995economic} and condom distribution \citep{zaric2001optimal,alistar2014hiv} and treatment programs like antiretroviral therapy \citep{lima2008expanded,alistar2014hiv} and methadone maintenance \citep{zaric2000hiv,zaric2000methadone}. However, our consideration of the effectiveness of recovery-improving interventions is different from that of the treatment effectiveness in \citep{alistar2014hiv} as the later does not increase the recovery rate but reduces the infectivity of susceptibles after treatment. It is to be noted that if both $\beta_{e}$ and $\gamma_{e}$ are set to zero, the model~\eqref{model_with_intervention} will be reduced to the original model~\eqref{basic_model}. Now in the presence of control strategy, we define control reproduction number $\mathcal{R}_{0}^{c}$ as
\begin{eqnarray*}
    \displaystyle\mathcal{R}_0^c &=& \displaystyle \frac{F_{\beta_{0}}(u)}{F_{\gamma_{0}}(1-u)+\mu}\\
    &=& \frac{\beta_0}{(1+\beta_e u)(\gamma_0(1+\gamma_e(1-u))+\mu)}.
\end{eqnarray*}

We use the phrases \textit{improving health care facilities} and \textit{improving recovery}, and also \textit{transmission-reducing} and \textit{prevention} programs interchangeably throughout the paper. 

\subsection{Model parametrization and simulations}
We take the numerical value of the model parameters like birth/death rate, disease transmission rate, and recovery rate from previous studies. The parameters related to effectiveness are varied in our study. The initial conditions, $S_0$ and $I_0$ are chosen with the assumption that initially, almost everyone in the population is susceptible. All simulations in the paper are done using parameters in Table~\ref{parameters_table} unless specified otherwise.

To obtain the optimal fraction of resources numerically, first, the values of $u$ are spaced linearly between $0$ to $1$ with suitable step-length and for each value of $u$, we integrate the model~\eqref{model_with_intervention} numerically using \textit{ode45} in MATLAB (Mathworks, R2018a) and calculate the objective function. We select that value of $u$ as optimal for which the objective function gives the minimum value.


  \begin{table}
  \tiny
  	\tabcolsep 7pt
  	\centering
  	\begin{tabular}{p{1cm} p{4.0cm} p{1.3cm} p{3cm}}
  		\hline
  		\hline\\
  		\textbf{Parameters} &\centering\textbf{Description}  &\centering\textbf{Value} & \textbf{References} \\
  		[2ex]
  		\hline\\
  		\centering$\mu$ & Natural birth and death rate of human  & $\frac{1}{70}$ $\rm year^{-1}$ & \citep{keeling2011modeling, martcheva2015introduction} \\[5ex]
  		\centering$\beta_0$ & Transmission rate & 0.8 $\rm day^{-1}$  & \citep{keeling2011modeling}\\[4ex]

  		\centering$\gamma_0$ & Recovery rate & 0.1 $\rm day^{-1}$ & \citep{keeling2011modeling,deka2019game}\\[4ex]

  		\centering$\beta_e$ & Effectiveness of transmission-reducing interventions & Varied &  - \\[5ex]
  		
  		\centering$\gamma_e$ & Effectiveness of recovery-improving interventions &  Varied & - \\[5ex]
  		
  		\centering$S_0$, $I_0$ & Initial proportion of susceptible, infected individuals & 0.999, 0.001& Assumed \\ [5ex]

  		\hline
  	\end{tabular}
  	\caption{\rm Description of parameters for the model~(\ref{basic_model}).}
  	\label{parameters_table}
  \end{table}
  

\section{Results}
\label{section_results}

We investigate the problem of optimal resource allocation in two scenarios: $(i)$ long-term or endemic dynamics of the disease, and $(ii)$ dynamics of the disease during an outbreak. In the first scenario, the demographic factors are considered, i.e., the birth and death rates are non-zero and in the second scenario, we ignore birth and natural death rates to capture the short-term dynamics of the disease.



{In the following subsections, we first describe the results for both of the long-term and outbreak scenarios for our model. Subsequently, we also analyze the model considering all three types of returns to scale for each of $F_{\beta_0}$ and $F_{\gamma_0}$ and all their possible combinations.}

\subsection{Resource allocation for long-term dynamics of the disease}\label{section-3(a)}
   To control long-term disease persistence, we take the proportion of infected individuals at equilibrium ($I^{*}$) as the objective function which is to be minimized in the presence of allocated resources. From the resource implemented model (\ref{model_with_intervention}), we can find the proportion of susceptible ($S^*$) and infected ($I^*$)  individuals at equilibrium analytically using null-isoclines as 

{\begin{equation*}
\displaystyle S^*=\begin{cases}
          1/\mathcal{R}_0^c \quad &\text{if} \,~ \mathcal{R}_0^c > 1 \\
          1 \quad &\text{if} \, ~\mathcal{R}_0^c\leq 1 \\
     \end{cases}
\end{equation*}
\begin{equation}
\displaystyle I^*=\begin{cases}
         \displaystyle \frac{\mu}{F_{\beta_{0}}}(\mathcal{R}_0^c-1) \quad &\text{if} \,~ \mathcal{R}_0^c > 1 \\
         0 \quad &\text{if} \, ~\mathcal{R}_0^c\leq 1 \\
     \end{cases}
     \label{I_last}
\end{equation}}
\noindent{When $\displaystyle \mathcal{R}_0^c > 1$, the endemic equilibrium is locally asymptotically stable while the disease-free equilibrium exists but is unstable. The equilibria merge when $\displaystyle \mathcal{R}_0^c = 1$, and only a stable disease-free equilibrium exists when $\displaystyle \mathcal{R}_0^c < 1$. 
This implies that the number of infected individuals asymptotically approaches a non-zero endemic level when $\displaystyle \mathcal{R}_0^c > 1$ and zero when $\displaystyle \mathcal{R}_0^c \leq 1$}. 

To find the optimal fraction of resources in the former case, i.e, when $\displaystyle \mathcal{R}_0^c > 1$, we differentiate $I^*$ with respect to $u$ and set it equal to zero. {Then simple algebraic manipulation allows us to solve for $u$, which we denote as $u_{\rm l}$:}

 \begin{equation}
 \displaystyle u_{\rm l} = \displaystyle\frac{\beta_e (\gamma_0+\gamma_0\gamma_e+\mu) -\sqrt{ \beta_0\beta_e\gamma_0\gamma_e}}{\beta_e\gamma_0 \gamma_e}.
 \label{u}
 \end{equation}
It is important to note that the fraction of resources allocated must be constrained between 0 and 1. {Taking this into account, it can be shown that when $\displaystyle \mathcal{R}_0^c \leq 1$, the optimal fraction of resources can lie anywhere in a certain interval $[u_{\rm l_1}, u_{\rm l_2}]\subseteq [0, 1]$ (see \ref{Appendix-A}). Hence, one can define the optimal fraction of resources to prevent transmission in case of long-term disease dynamics, $u^*_{\rm long}$, as follows: }
\begin{equation}
{u^*_{\rm long}=\begin{cases}
\left.\begin{aligned}
0\qquad\qquad &\text{if} \,~ u_{\rm l} \leq0\\
u_{\rm l}\qquad\qquad &\text{if} \, ~0< u_{\rm l}< 1\\
1\qquad\qquad &\text{if} \, ~u_{\rm l} \geq1\\
\end{aligned}\right\}~\mathcal{R}_0^c> 1\\
\left.\begin{aligned}
 \displaystyle \big[u_{\rm l_1}, u_{\rm l_2}\big]\hspace{10pt} \subseteq [0, 1]\hspace{30.5pt}
\end{aligned}\right\}~\mathcal{R}_0^c\leq 1\\
     \end{cases}}
     \label{u_long}
\end{equation}

\begin{figure}[H]
    	\begin{center}
    	\includegraphics[width=1\textwidth]{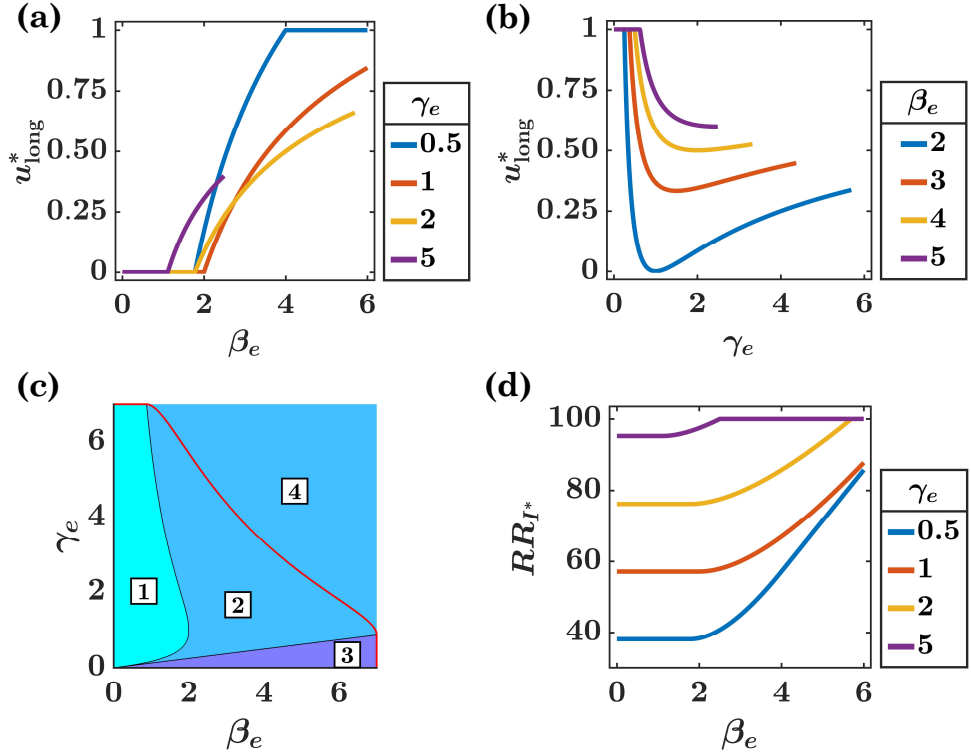}
    	\end{center}
    	\caption{\rm Role of effectiveness parameters in determining the optimal fraction of allocated resources to control disease dynamics in long-term scenario. Optimal fraction of resources (a) with respect to $\beta_{e}$, (b) with respect to $\gamma_{e}$, (c) when both the effectiveness parameters,  $\beta_{e}$ and $\gamma_{e}$, are varied simultaneously. The red line in (c) represents $\mathcal{R}_0^c=1$. Here, unique $u^*_{\rm long}$ is associated with $\displaystyle \mathcal{R}_0^c > 1$ (displayed as $\boxed{1}$, $\boxed{2}$ and \boxed{3}), while non-unique $u^*_{\rm long}$ is associated with $\displaystyle \mathcal{R}_0^c \leq 1$ (displayed as $\boxed{4}$). {$\boxed{1}$ represents the region where $u^*_{\rm long}=0$, i.e., allocating entire resources toward the recovery-improvement programs is the optimal strategy. $\boxed{3}$ represents the region where the allocation of entire resources towards transmission-reducing programs is the optimal strategy (i.e., $u^*_{\rm long}=1$). $\boxed{2}$ and $\boxed{4}$ represent the regions for resource sharing between two types of intervention programs. $u^*_{\rm long}$ has a unique value between 0 and 1 in $\boxed{2}$; $u^*_{\rm long}$ lies within an interval in $\boxed{4}$. (d) Relative reduction of $I^*$ with respect to $\beta_e$ for different values of $\gamma_{e}$.}}
    	\label{long-term_figure}
    \end{figure}

Now, we explore the behavior of $u^*_{\rm long}$ with respect to the effectiveness of intervention programs $\beta_e$, $\gamma_e$ when $\displaystyle \mathcal{R}_0^c > 1$. Here, $\displaystyle\lim_{\beta_e \to 0 } u^*_{\rm long}=0$ (as $\displaystyle \lim_{\beta_e \to 0 } u_{\rm l} \xrightarrow{} -\infty$) indicates that when $\beta_e$ is too low the optimal strategy is to allocate the entire resources to recovery improvement programs. {This holds true until a threshold, $\displaystyle\beta_e^l=\frac{\beta_0\gamma_0\gamma_e}{(\gamma_0+\gamma_0\gamma_e+\mu)^2}$ above which $u^*_{\rm long}$ increases with $\beta_e$ implying the need to allocate increasing fractions of resources to prevention programs (see Fig.~\ref{long-term_figure}(a)). It is interesting to note that $\beta_e^l$ is non-monotonic in $\gamma_e$. As a result, we can see that the threshold for $\gamma_e=1$ shifts towards right from the threshold for $\gamma_e=0.5$, but for higher values of $\gamma_e$ ($\gamma_e$=2, 5), the threshold shifts toward left (see Fig.~\ref{long-term_figure}(a)).} 


Similarly, $\displaystyle\lim_{\gamma_e \to 0 } u^*_{\rm long}=1$ (as $\displaystyle\lim_{\gamma_e \to 0 } u_{\rm l} \xrightarrow{} +\infty$) implies that when $\gamma_e$ is too low, the optimal strategy is to devote entire resources to transmission-reducing programs. {Again, this holds true only until a threshold $\displaystyle \gamma_e^l=\frac{\beta_e(\gamma_0+\mu)^2}{\beta_0\gamma_0}$ above which $u^*_{\rm long}$ starts to decrease with $\gamma_e$.} This trend persists until a second threshold, $\displaystyle\widetilde{\gamma_e}=\frac{4\beta_e(\gamma_0+\mu)^2}{\beta_0\gamma_0}$ after which the optimal strategy is to allocate decreasing fractions of resources toward treatment programs (see Fig.~\ref{long-term_figure}(b)). {Furthermore, the sensitivity of our findings to the baseline model parameters ($\beta_0$, $\gamma_0$, and $\mu$) is discussed in~\ref{Appendix-A} and displayed in Fig.~\ref{sensitivity_figure}.}

In order to better understand the role of the effectiveness parameters in optimal resource allocation, the two-parameter space ($\beta_e-\gamma_e$) can be divided into four regions (see Fig.~\ref{long-term_figure}(c)). {We vary $\beta_e$ and $\gamma_e$ till the values such that only one type of intervention alone cannot achieve an asymptotically stable disease-free state.} Here \boxed{1}, \boxed{2}, and \boxed{3} are associated with the {locally asymptotically stable endemic state of the prevalence} (where $\displaystyle \mathcal{R}_0^{\rm c} > 1$). In \boxed{1} and \boxed{3}, the optimal strategy is to allocate entire resources to recovery improvement (i.e., $u^*_{\rm long}$ = 0) and transmission-reducing programs (i.e., $u^*_{\rm long}$ = 1) respectively. In \boxed{2}, there will be unique optimal strategies between 0 and 1 determined by the effectiveness parameters which signifies the need to allocate resources to both prevention and treatment. The resulting control reproduction number can then be expressed as:
\begin{equation}
 \mathcal{R}_0^{\rm long} = \displaystyle \frac{\sqrt{\beta_0\beta_e\gamma_0\gamma_e}}{\beta_e(\gamma_0+\mu)+\gamma_0\gamma_e(1+\beta_e)-\sqrt{\beta_0\beta_e\gamma_0\gamma_e}}.
   \label{R_0_long-term}
 \end{equation} 
 
 Here, the lines separating \boxed{2} from \boxed{1} and \boxed{3} is formed by the threshold values $\beta_e^l$ and $\gamma_e^l$ mentioned above.
\boxed{4} corresponds to $\mathcal{R}_0^c\leq1$ where $u^*_{\rm long}$ is not unique but can vary between certain ranges (see expression~\eqref{u_long_interval} and Fig.~\ref{unique_non_unique_u_long_figure} in \ref{Appendix-A}). { With increasing $\beta_e$ and $\gamma_e$, the range of values of $u^*_{\rm long}$ in \boxed{4} increases, signifying a greater flexibility of choice of optimal strategy (see  Fig.~\ref{unique_non_unique_u_long_figure}, \ref{Appendix-A}).} The line $\displaystyle\mathcal{R}_0^{\rm c} = 1$ denotes the borderline between \boxed{4} and \boxed{2} ({red line in Fig.~\ref{long-term_figure}(c)}). It can also be interpreted as the line of critical thresholds $\displaystyle \beta_e^{\rm crit}(\gamma_e)$ or $\displaystyle \gamma_e^{\rm crit}(\beta_e)$, i.e., the least values of $\beta_e$ and $\gamma_e$ which will lead to stable disease-free state asymptotically under optimal resource allocation. Overall, we see that the optimal strategy {changes from allocating resources in a single intervention in \boxed{1} and \boxed{3} to dividing resources between both interventions in \boxed{2} and \boxed{4}} under different effectiveness parameters.

Now we quantify the relative impact of resource allocation on disease dynamics. For this, we consider the uncontrolled epidemic (i.e., when both $\beta_{e}$ and $\gamma_{e}$ are zero) as the baseline scenario. We then calculate the percentage of relative reduction ($RR_{I^*}$) of the objective functions $I^*$ by varying the effectiveness parameters in presence of optimal resources (see Fig.~\ref{long-term_figure}(d)). For small values of $\gamma_{e}$, such as $\gamma_{e}=0.5, 1, 2$, if we vary $\beta_{e}$ in the interval $0\leq \beta_{e} \lessapprox 2$, the relative reduction in $I^{*}$  by allocating the available resources optimally stays almost same with approximately $38\%$, $57\%$, $76\%$ respectively (see Fig.~\ref{long-term_figure}(d)). Since, for these $\gamma_{e}$ values within $0\leq \beta_{e}\lessapprox 2$, $u_{\rm long}^{*}=0$ is found to be optimal (see Fig.~\ref{long-term_figure}(a)), therefore we do not observe any role of $\beta_{e}$ on the reduction of $I^{*}$. However, if we increase $\beta_e$ further ($2\leq \beta_{e} \leq 6$), we see gradual increase in $RR_{I^{*}}$ (see Fig.~\ref{long-term_figure}(d)). This is because $u_{\rm long}^{*}$ is non-zero in this case and both $\beta_e$, $\gamma_{e}$ have certain contributions. If $\gamma_{e}$ is fixed to a relatively higher value, such as $\gamma_{e} = 5$ , $RR_{I^*}$ can reach about $95\%$ (see Fig.~\ref{long-term_figure}(d)).

\subsection{Resource allocation for an outbreak scenario}
For an outbreak scenario, we ignore natural birth and death in our model, i.e., we fix $\mu=0$ in our model \eqref{model_with_intervention}. In this case, the control reproduction number can be expressed as $ \displaystyle \mathcal{R}_0^{\rm out}$ = $\displaystyle \frac{\beta_0}{\gamma_0(1+\beta_e u)(1+\gamma_e(1-u))}$. {One can infer about the basic characteristics of an outbreak from the sign of $\displaystyle \frac{dI}{dt}\mid_{t=0}$ or equivalently $1-S_0\mathcal{R}_0^{\rm out}$. When  $ S_0\mathcal{R}_0^{\rm out}>1$, the number of infected individuals initially increases until it reaches its maximum and subsequently decays. When $\displaystyle S_0\mathcal{R}_0^{\rm out}=1$, the outbreak is initially in a stationary state and subsequently decays, and for $\displaystyle S_0\mathcal{R}_0^{\rm out}<1$, the outbreak is initially in a decaying state and remains to decay for all subsequent time. The maximum number of infected individuals at any one time, attained during the outbreak is defined as the peak of the epidemic.} Since the peak is a primary indicator of the severity of an outbreak, lowering it is one of the primary goals for reducing the burden of outbreaks.

For this purpose, we derive the epidemic peak ($I_{\rm max}$) analytically which can be expressed as follows (see \ref{Appendix-B})
{ \begin{equation}
\displaystyle I_{\rm max}=\begin{cases}
         \displaystyle S_0+I_0-\frac{\bigg(\ln{\Big(S_0\mathcal{R}_0^{\rm out}\Big)}+1\bigg)}{\mathcal{R}_0^{\rm out}}\quad &\text{if} \,~ S_0\mathcal{R}_0^{\rm out} > 1 \\
         I_0 \quad &\text{if} \, ~S_0\mathcal{R}_0^{\rm out}\leq 1 \\
     \end{cases}
     \label{I_max}
\end{equation}}

In case of initial growth of outbreak size (i.e., $S_0\mathcal{R}_0^{\rm out} > 1$), the optimal fraction of resources can be obtained by differentiating $I_{\rm max}$ with respect to $u$, setting it equal to zero and solving algebraically (see~\ref{Appendix-B}):
\begin{equation}
 u_{\rm p} =\displaystyle\frac{1}{2}\Big(1-\frac{1}{\beta_e}+\frac{1}{\gamma_e}\Big) 
   \label{u_p}
\end{equation}
{ Further, when the outbreak is initially in a stationary or decaying state, one can solve the inequality $S_0\mathcal{R}_0^{\rm out}\leq1$ for $u$ to obtain the optimal fraction of resources in the form of an interval, $[u_{\rm p_1}, u_{\rm p_2}]$ (see~\ref{Appendix-B}).} As the fraction of allocated resources must be constrained between 0 and 1, a new quantity $u^*_{\rm peak}$ is defined as the optimal fraction of resources invested in transmission-reducing programs such that (see~\ref{Appendix-B} for details)
\begin{equation}
{u^*_{\rm peak}=\begin{cases}
\left.\begin{aligned}
0\qquad\qquad &\text{if} \,~ u_{\rm p} \leq0\\
u_{\rm p}\qquad\qquad &\text{if} \, ~0< u_{\rm p}< 1\\
1\qquad\qquad &\text{if} \, ~u_{\rm p} \geq1\\
\end{aligned}\right\}~S_0\mathcal{R}_0^{\rm out}> 1\\
\left.\begin{aligned}
 \displaystyle \big[u_{\rm p_1}, u_{\rm p_2}\big]\hspace{8pt}\subseteq [0, 1]\hspace{32.5pt}
\end{aligned}\right\}~S_0\mathcal{R}_0^{\rm out}\leq 1\\
     \end{cases}}
     \label{u_peak}
\end{equation}

\begin{figure}[H]
    	\begin{center}
    		\includegraphics[width=1\textwidth]{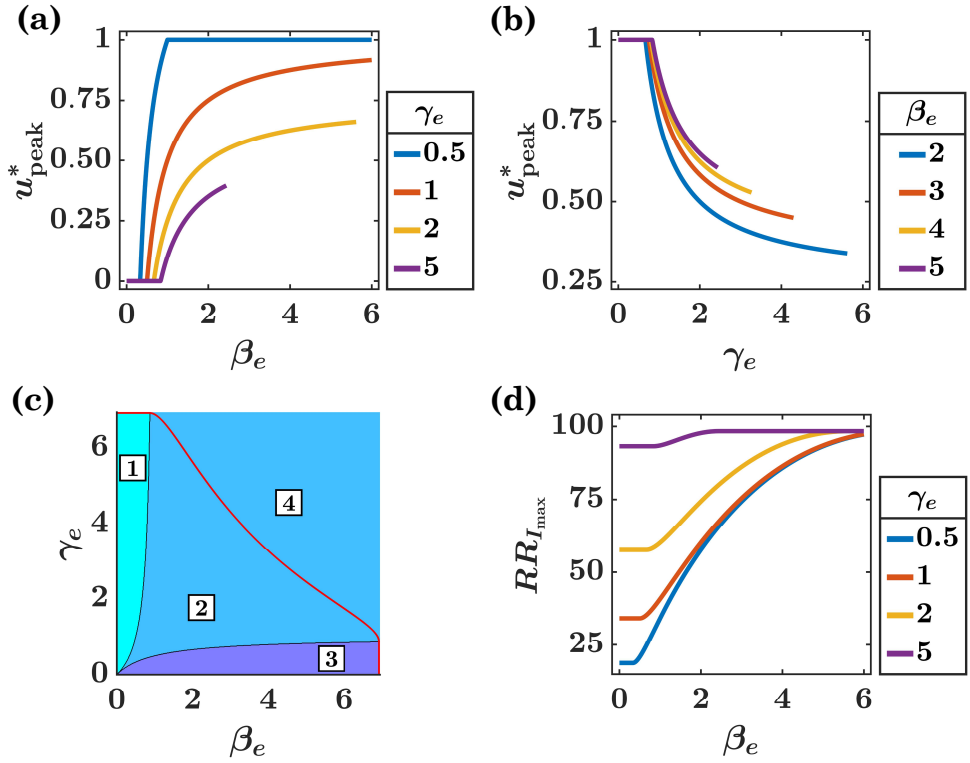}
    	\end{center}
    	\caption{\rm Role of effectiveness parameters in determining the optimal fraction of allocated resources to control disease in an outbreak scenario. Optimal fraction of resources (a) with respect to $\beta_{e}$, (b) with respect to $\gamma_{e}$, (c) when both the effectiveness parameters, $\beta_{e}$ and $\gamma_{e}$, are varied simultaneously. The red line represents $S_0\mathcal{R}_0^{\rm out}=1$. Unique $u^*_{\rm peak}$ corresponds to the initial growth of the outbreak size (i.e., $\displaystyle S_0\mathcal{R}_0^{\rm out} > 1$; displayed as $\boxed{1}$, $\boxed{2}$ and \boxed{3}), while non-unique $u^*_{\rm peak}$ is associated with the initial decay of the outbreak size (i.e., $\displaystyle S_0\mathcal{R}_0^{\rm out} < 1$; displayed as $\boxed{4}$. {$\boxed{1}$ represents the region where $u^*_{\rm peak}=0$, i.e., allocating entire resources toward the recovery-improvement programs is the optimal strategy. $\boxed{3}$ represents the region where the allocation of entire resources towards transmission-reducing programs is the optimal strategy (i.e., $u^*_{\rm peak}=1$). $\boxed{2}$ and $\boxed{4}$ represent the regions for resource sharing between two types of intervention programs. $u^*_{\rm peak}$ has a unique value between 0 and 1 in $\boxed{2}$; $u^*_{\rm peak}$ lies within an interval in $\boxed{4}$. (d) Relative reduction of $I_{\rm max}$ with respect to $\beta_e$ for different values of $\gamma_{e}$.}}
    
    	\label{outbreak_figure}
    \end{figure}

It can be observed from the Eq.~\eqref{u_p} of $u_{\rm p}$ that $u^*_{\rm peak}$ depends only on the effectiveness of intervention programs $\beta_e$, $\gamma_e$ when $S_0\mathcal{R}_0^{\rm out}> 1$ and is independent of the transmission rate and recovery rate $\beta_0$, $\gamma_0$. Similar to the long-term scenario, we explore the behavior of $u^*_{\rm peak}$ with respect to $\beta_e$ and $\gamma_e$. Here, $\displaystyle\lim_{\beta_e \to 0 } u^*_{\rm peak}=0$ (as $\displaystyle \lim_{\beta_e \to 0 } u_{\rm p} \xrightarrow{} -\infty$) indicates that when $\beta_e$ is too low, the optimal strategy is to allocate entire resources to recovery improvement programs. {This holds true until a certain threshold, $\displaystyle\beta_e^p=\frac{\gamma_e}{1+\gamma_e}$ after which $u^*_{\rm peak}$ increases with $\beta_e$. (Fig.~\ref{outbreak_figure}(a)).} Likewise, $\displaystyle\lim_{\gamma_e \to 0 } u^*_{\rm peak}=1$ (as $\displaystyle\lim_{\gamma_e \to 0 } u_{\rm p} \xrightarrow{} +\infty$) implies when $\gamma_e$ is too low, the optimal strategy is to devote entire resources to transmission-reducing programs. {Again, this holds true only until a threshold $\displaystyle \gamma_e^p=\frac{\beta_e}{1+\beta_e}$ above which $ u^*_{\rm peak}$ starts to decrease with $\gamma_e$ (see Fig.~\ref{outbreak_figure}(b)).
Notably, in this case, the monotonic behaviour of $u^*_{\rm peak}$ with respect to $\beta_e$ and $\gamma_e$ indicates that} the optimal control strategy is to allocate more resources towards intervention programs with better effectiveness (Fig.~\ref{outbreak_figure}((a)(b)).



Parameter space ($\beta_e-\gamma_e$) can be divided into four regions on the basis of optimal strategy (see Fig.~\ref{outbreak_figure}(c)). { 
 Similar to Fig.~\ref{long-term_figure}(c), $\beta_e$ and $\gamma_e$ are varied till the values such that resource allocation to only one type of intervention cannot satisfy the condition for the initial decay state of the outbreak.} \boxed{1}, \boxed{2} and \boxed{3} {correspond to $\displaystyle S_0\mathcal{R}_0^{\rm out}>1$}, where the optimal fraction of resources towards transmission prevention, $u^*_{\rm peak}$ is given by  $0, u_{\rm p}$ and $1$ respectively. {In \boxed{2}, where $u_{\rm p}$ is the unique optimal strategy, the control reproduction number for the outbreak scenario is given by:

 \begin{equation}
  \displaystyle \mathcal{R}_0^{\rm peak} = \displaystyle \frac{4\beta_0\beta_e\gamma_e}{\gamma_0(\beta_e\gamma_e+\beta_e+\gamma_e)^2}.
   \label{R_0_peak}
 \end{equation}
 
The curves that separate \boxed{1} and \boxed{3} from \boxed{2} correspond to the threshold values  $\beta_e^p$ and $\gamma_e^p$ respectively. In contrary to Fig.~\ref{long-term_figure}(c), here, \boxed{1} and \boxed{3} are symmetric about the line $\beta_e=\gamma_e$. This can be easily explained by the fact that both $\beta_e^p(\gamma_e)$ and $\gamma_e^p(\beta_e)$ satisfy the same underlying function.} {Further, it is interesting to note that $\displaystyle u_{\rm p}(\beta_e, \gamma_e) = 1- u_{\rm p}(\gamma_e, \beta_e)$. This implies, if the effectiveness of prevention and treatment programs are reversed in case of certain diseases, then the optimal strategy is corresponding reversal of resource allocation.} On optimal utilization of resources with increased effectiveness parameters, {the outbreak is unable to initially grow $(\displaystyle S_0\mathcal{R}_0^{\rm out}\leq1$), which is represented by \boxed{4} (see~\ref{Appendix-B}).} Within \boxed{4}, $u^*_{\rm peak}$ is non-unique and varies between certain ranges (see expression~\eqref{u_peak_interval} in \ref{Appendix-B}). The line $\displaystyle S_0\mathcal{R}_0^{\rm out} = 1$ ({red line}) separates \boxed{4} from \boxed{2} in Fig.~\ref{outbreak_figure}(c). It can also be interpreted as the line of critical values $\displaystyle \beta_e^{\rm crit}(\gamma_e)$ or $\displaystyle \gamma_e^{\rm crit}(\beta_e)$, i.e., least values of $\beta_e$ and $\gamma_e$ which are sufficient to prevent the outbreak from growing initially under optimal resource allocation.

With respect to the baseline scenario of uncontrolled epidemics ($\beta_e, \gamma_e=0$), we calculate the percentage of relative reduction ($RR_{I_{\rm max}}$) of $I_{\rm max}$ by varying the effectiveness parameters in presence of optimal resources (see Fig.~\ref{outbreak_figure}(d)). For lower values of $\beta_{e}$, when $u^{*}_{\rm peak}=0$ is optimal (see Fig.~\ref{outbreak_figure}(a)), we see from Fig.~\ref{outbreak_figure}(d) that approximately $18\%$, $34\%$, $57\%$ and $93\%$ reduction can be achieved for $\gamma_{e}$=0.5, 1, 2 and 5 respectively. Further increase in $\beta_e$ leads to gradual increase in $RR_{I_{\rm max}}$.

\subsubsection*{Impact of delays in resource implementation}
\begin{figure}[H]
    	\begin{center}
     \includegraphics[width=0.9\textwidth]{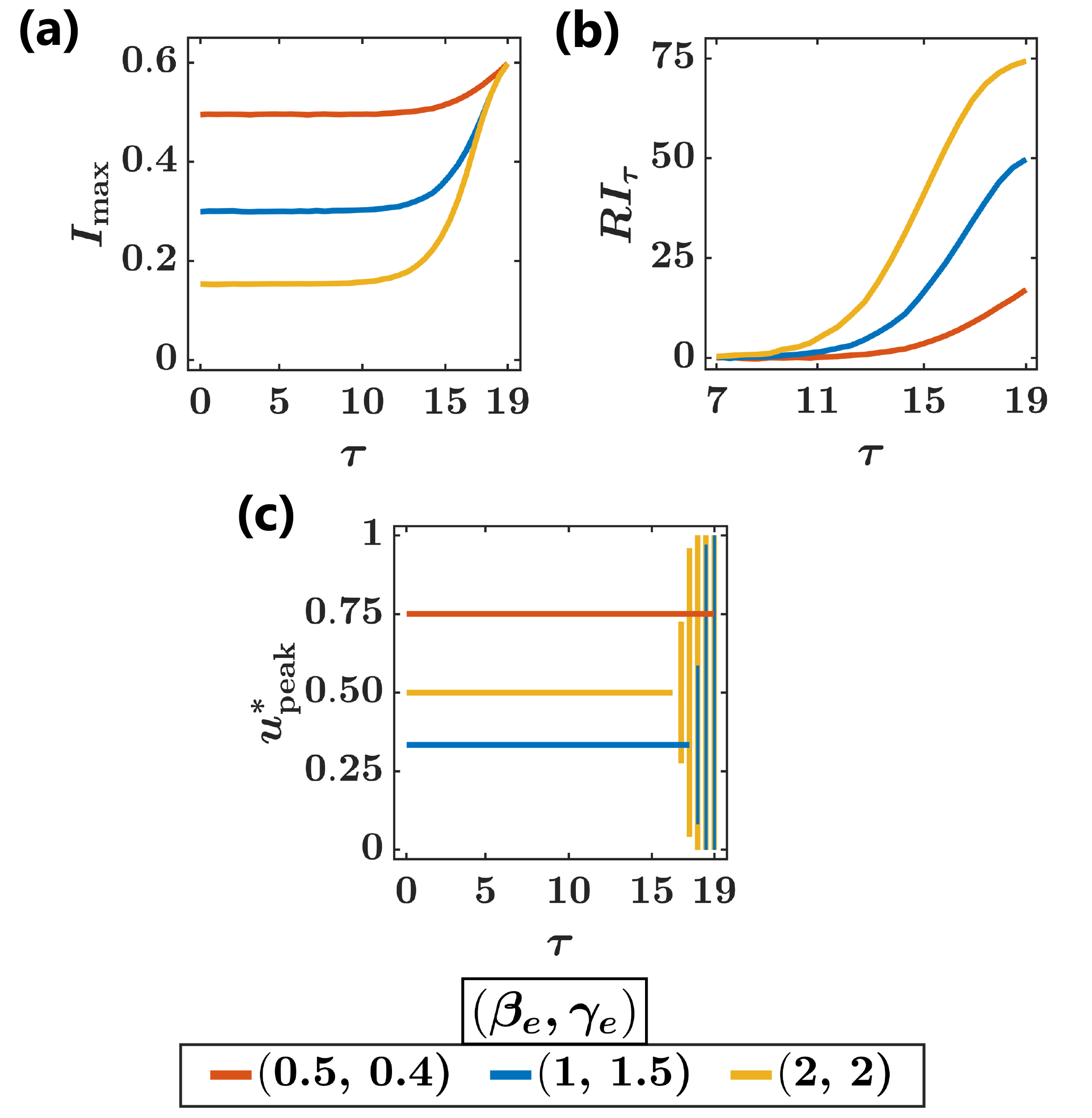}
    	\end{center}
    	\caption{\rm Effect of delay ($\tau$) in implementing interventions during a disease outbreak for different combinations of $\beta_{e}$ and $\gamma_{e}$. (a) The peak of infection ($I_{\rm max}$) as a function of $\tau$. (b) Percentage of the relative increase of $I_{\rm max}$ compared to immediate intervention (i.e., $\tau=0$) with respect to $\tau$. (c) The optimal fraction of resources ($u^*_{\rm peak}$) aiming minimization of the peak of infection as a function of delay $\tau$. The vertical lines represent the range of possible values of $u^*_{\rm peak}$ when non-unique.}
    	\label{delay_figure}
    \end{figure}

In reality, it is quite reasonable to expect a delay between the implementation of intervention measures and the onset of an outbreak due to inappropriate information about it at an early stage. {To explore the impact of such delays during outbreaks, we calculate the peak of infection ($I_{\rm max}$) when the control measures are implemented $\tau$ days after the onset of the outbreak (see Fig.~\ref{delay_figure}(a)). We also calculate the percentage of relative increase ($RI_{\tau}$) in the peak of infection for each delay time, $\tau$, compared to the immediate implementation ($\tau=0$) (see Fig.~\ref{delay_figure}(b)). Lastly, we calculate the corresponding optimal fraction of resources ($u^{*}_{\rm peak}$) as a function of delay time, $\tau$, for different values of effectiveness parameters in Fig.~\ref{delay_figure}(c).} In our model, for an uncontrolled epidemic (i.e., $\beta_e = \gamma_e = 0$), the infection curve achieves its peak around $t=19$ days. Since we are interested in peak minimization (lowering the curve of infection), we vary the implementation delay parameter ($\tau$) from days $t=0$ to $t=19$.

For lower values of both the effectiveness parameters $\beta_{e}$ and $\gamma_{e}$ (for instance, $\beta_e=0.5$ and $\gamma_e=0.4$), the peak of infection ($I_{\rm max}$) remains almost the same until day $\tau=12$ (see Fig.~\ref{delay_figure}(a), red curve). That means both interventions with lower effectiveness allow a delay of up to 12 days in allocating resources without any significant consequences in terms of the peak of infection. However, if the allocation of resources is delayed further, a gradual increase in the peak of infection can be observed. We find that after a delay of $12$ days, $RI_{\tau}$ increases gradually and reaches about $17\%$ at day $\tau=19$ (see Fig.~\ref{delay_figure}(b), red curve) for lower effectiveness parameters. On the other hand, for higher values of both effectiveness parameters (for instance, $\beta_e=2$ and $\gamma_e=2$), if the implementation is delayed more than 7 days, a larger increase (approximately $74\%$ at $\tau=19$;  Fig.~\ref{delay_figure}(a)(b), yellow curve) in $RI_{\tau}$ is observed compared to the lower effectiveness case. This is because interventions with higher effectiveness are capable of greater reductions in the peak of infection and delays in implementing these interventions lead to a significant relative increase in the same. It is also to be noted that for lower values of effectiveness parameters ($\beta_{e}=0.5$ and $\gamma_{e}=0.4$), the optimal fraction of resources ($u^*_{\rm peak}$) remains unchanged with respect to delay (see Fig.~\ref{delay_figure}(c), red curve). This is because $u^*_{\rm peak}$ depends only on $\beta_e$ and $\gamma_e$ (see Eq.~\eqref{u_peak}). Also, for lower values of these effectiveness parameters, the condition, $\displaystyle S_{\tau}\mathcal{R}_0^{\rm out}>1$ is satisfied for all $\tau$ resulting in a unique optimal value (where $S_{\tau}$ denotes the proportion of susceptible individuals for the uncontrolled epidemic at $t=\tau$ day after the onset of the outbreak). For higher values of $\beta_{e}$ and $\gamma_{e}$, $u^*_{\rm peak}$ behaves in a slightly different way. Similar to the above-mentioned case, $u^*_{\rm peak}$ stays the same and takes a unique value until a threshold of $\tau$ is reached, where the quantity, $\displaystyle S_{\tau}\mathcal{R}_0^{\rm out}$ becomes less or equals to 1 (as susceptible population depletes over time). Consequently, $u^*_{\rm peak}$ becomes non-unique and can take a range of possible values from an interval (see Fig.~\ref{delay_figure}(c), blue and yellow curves).

\begin{figure}[H]
    	\begin{center}
    	\includegraphics[width=1\textwidth]{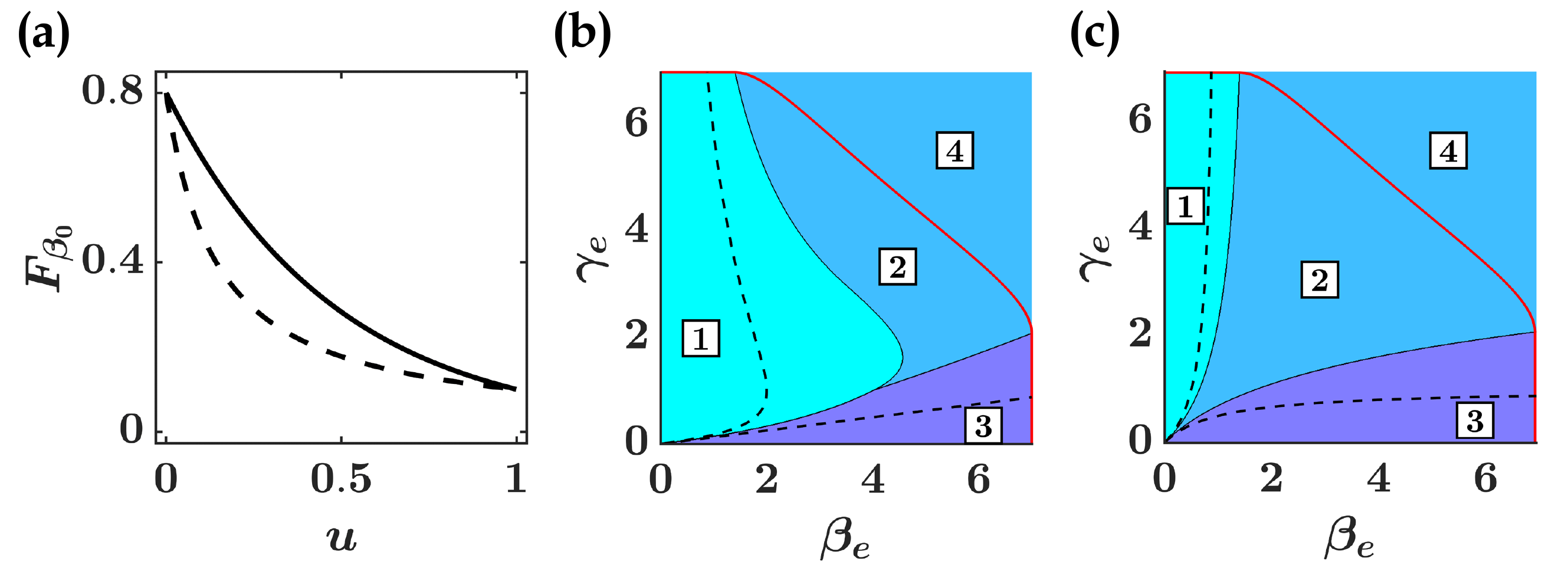}
    	\end{center}
    	\caption{{\rm (a) The solid and the dashed lines represent the production functions $\displaystyle F_{\beta_0}=\displaystyle \beta_0 e^{-u\ln{(1+\beta_e)}}$ and $\displaystyle F_{\beta_0}=\displaystyle \frac{\beta_0}{1+\beta_e u}$ respectively. Role of effectiveness parameters in the optimal fraction of allocated resources to control disease dynamics in $(a)$ long-term and $(b)$ outbreak scenario. Here the red lines, $\mathcal{R}_0^c=1$ and $S_0\mathcal{R}_0^{\rm out}=1$ in (b) and (c) respectively are shown only for  $\displaystyle F_{\beta_0}=\displaystyle \beta_0 e^{-u\ln{(1+\beta_e)}}$. An increase in \boxed{1}, \boxed{3} are observed compared to the original form in Fig.~\ref{long-term_figure}(c) and Fig.~\ref{outbreak_figure}(c).}}
   \label{robustness_figure}
    \end{figure}

\subsection{Dependence of resource allocation on production functions}

{ To study the robustness of our results, we choose a different $F_{\beta_0}$ with decreasing returns given as $F_{\beta_0} = \beta_0 e^{-u\ln{(1+\beta_e)}}$ (solid line in Fig.~\ref{robustness_figure}(a)). A similar exponential function has been used earlier to estimate the production functions for HIV prevention programs \citep{zaric2001optimal, alistar2014hiv}. For extreme values of $u$, the new $F_{\beta_0}$ is equivalent to that assumed in the previous sections (see Fig.~\ref{robustness_figure}(a)). As we want to keep the returns to scale type same for better comparison, and since $F_{\gamma_0}$ exhibits constant returns to scale, it is kept unchanged. This implies that $\mathcal{R}_0^c$, evaluated at the extreme values of $u$, are same as in Sec.~\ref{section-3(a)}. As a result, in the long-term scenario, the line $\mathcal{R}_0^c=1$, intersects $\beta_e$ and $\gamma_e$ axis at the same values as in Fig.~\ref{long-term_figure}(c), thereby facilitating comparison. For the outbreak scenario, similar arguments hold for the line, $S_0\mathcal{R}_0^{\rm out}=1$.

We calculate optimal strategies for both long-term and outbreak scenarios for different effectiveness parameters in Fig.~\ref{robustness_figure}(b),(c) and compare \boxed{1} - \boxed{4} with that of Fig.~\ref{long-term_figure}(c) and Fig.~\ref{outbreak_figure}(c). For the long-term case, the non-monotonic behaviour of the $\beta_e^l$ is retained, which suggests that this property might be independent of the functional forms of $F_{\beta_0}$ with decreasing returns to scale. On the other hand, for the outbreak scenario, the symmetric nature of \boxed{1} and \boxed{3} about the line $\beta_e= \gamma_e$ is lost which implies that it can only be true in case of certain production functions. Further, in this case, one can easily note the increase in the area occupied by \boxed{1} and \boxed{3}. This signifies a greater possibility that allocating resources to either one of the prevention or treatment programs is the optimal strategy.}

\begin{figure}[H]
    	\begin{center}
    		 \includegraphics[width=1\textwidth]{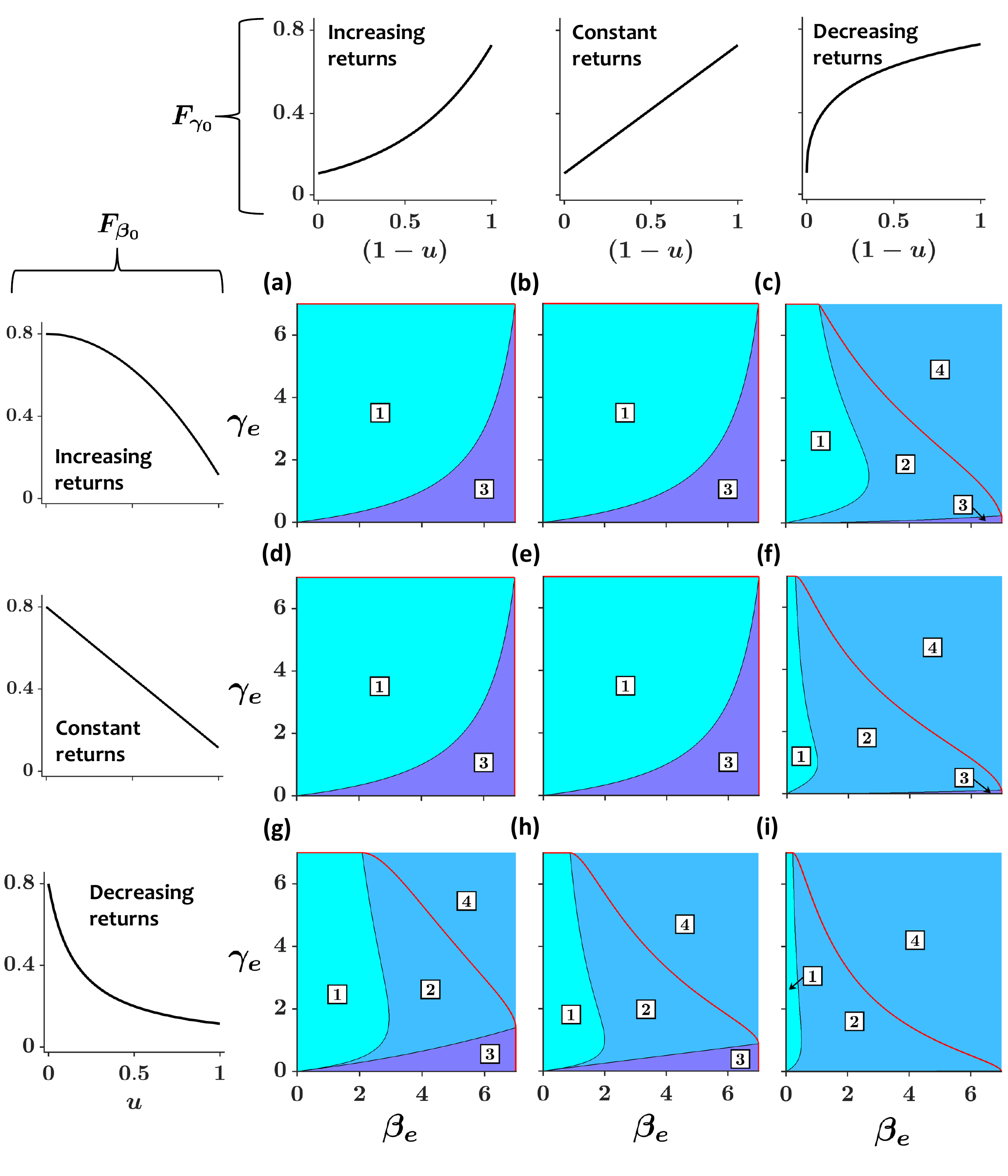}
    	\end{center}
    	\caption{{\rm Comparison of the results considering three types of returns to scale for both $F_{\beta_0}$ and $F_{\gamma_0}$ to explore the dependence of optimal resource allocation on production functions for control of disease in long-term scenario. The characteristics of $u^*_{\rm long}$ within \boxed{1}, \boxed{2}, \boxed{3} and \boxed{4} are the same as described in Fig.~\ref{long-term_figure}(c). For specific forms of $F_{\beta_{0}}$ and $F_{\gamma_{0}}$, refer to \ref{Appendix-B}, Fig.~\ref{outbreak_9_figure} caption.
     }
    	}
    	\label{long-term_9_figure}
    \end{figure}

{ Till now, we considered only decreasing and constant returns to scale for transmission-reducing and recovery-improvement programs, respectively. However, it is important to understand the results obtained above in the context of the other returns types of production functions. Hence, in this section, we consider each of $F_{\beta_0}(u)$ and $F_{\gamma_0}(1-u)$ with increasing, constant and decreasing returns to scale and compare the results among all their combinations (see Fig.~\ref{long-term_9_figure}). For a more meaningful comparison between results, we consider that the three types of returns to scale for $\displaystyle F_{\beta_0}$ (same for $\displaystyle F_{\gamma_0}$) are equivalent to each other whenever $u = 0$ or $u = 1$ (see Fig.~\ref{production_function_figure}). The exact functional forms for $F_{\beta_0}(u)$ and $F_{\gamma_0}(1-u)$ used in Fig.~\ref{long-term_9_figure} are described in the caption of Fig.~\ref{outbreak_9_figure} in \ref{Appendix-B}.

We observe that when $F_{\beta_0}$ gives either increasing or constant returns to scale and the same holds for $F_{\gamma_0}$, the optimal fraction of resources, $u^{*}_{\rm long}$ takes only extreme values (either 0 in \boxed{1} or 1 in \boxed{3}) (see Fig.~\ref{long-term_9_figure} (a, b, d, e)). Interestingly, in these cases, $u^*_{\rm long}$ is independent of the functional forms of $F_{\beta_0}$, $F_{\gamma_0}$ and has same value for a particular $\beta_e$ and $\gamma_e$ (see Proposition.~\ref{proposition-1}, \ref{Appendix-A}). On the other hand, when either or both of $F_{\beta_0}$ and $F_{\gamma_0}$ having decreasing returns to scale, $u^{*}_{\rm long}$ also takes intermediate values between 0 and 1 (see Fig.~\ref{long-term_9_figure} (c, f-i). Further, if $F_{\beta_0}$ ($F_{\gamma_0}$) gives decreasing returns to scale and $F_{\gamma_0}$ ($F_{\beta_0}$) changes from increasing to decreasing returns to scale, we observe the region \boxed{1} and \boxed{3} decreases (see Fig.~\ref{long-term_9_figure} (c, f, i) and (g, h, i)). Moreover, when both intervention programs have decreasing returns to scale, a combined prevention and treatment program is most likely the optimal strategy (see Fig.~\ref{long-term_9_figure}(i)). These results indicate that decreasing returns from investments into the intervention programs promote the necessity for resource sharing. Additionally, moving from increasing to decreasing returns to scale, relatively lower effectiveness of interventions can be sufficient to obtain an asymptotically stable disease-free state, i.e., $\mathcal{R}_0^c<1$. For instance, when $\beta_e=\gamma_e=4$, in Fig.~\ref{long-term_9_figure}(g), we have $\mathcal{R}_0^c>1$ but the same conditions in Fig.~\ref{long-term_9_figure}(i) yield $\mathcal{R}_0^c<1$. The intuition behind such observation can be easily understood from the shape of the curves, $F_{\beta_0}$ and $F_{\gamma_0}$. When compared to increasing or constant returns, decreasing returns to scale for $F_{\beta_0}$, $F_{\gamma_0}$ implies that relatively lower transmission and higher recovery rates can be achieved for the same investment (except at $u=0,1$). All the results described above also hold true for the outbreak scenario (see Fig.~\ref{outbreak_9_figure} in \ref{Appendix-B}).} {It is important to mention that for the long-term scenario, the non-monotonic behaviour of $\beta_e^l$ with respect to $\gamma_e$ is observed when at least one of $F_{\beta_0}$ and $F_{\gamma_0}$ has decreasing returns to scale (see Fig.~\ref{long-term_9_figure} (c, f-i)).}

 \section{Discussion}
\label{sec:discussion}
Designing effective intervention strategies is of utmost importance in order to cope with disease outbreaks. Particularly, in low-income countries, resource-constrained situations make it challenging for authorities to allocate resources in an optimal manner to mitigate epidemic scenarios. In this context, the resource allocation trade-off between prevention of transmission and better medical treatment is of significance, yet not well understood. 
We developed a simple model based on the SIR framework which takes into account {production functions for intervention programs and associated effectiveness} to address this gap. The interplay between the effectiveness of the two major types of interventions and their implication on optimal resource allocation has been explored {under two common epidemic scenarios, viz., long-term and outbreak. While the former case concerns the minimization of the proportion of infected individuals at equilibrium, the primary objective in the later is to minimize the peak of infection during a disease outbreak.} For both scenarios, we observe the transitions from allocation of entire resources to only one intervention type to optimal sharing of resources between both types of interventions under different effectiveness parameters settings (Fig.~\ref{long-term_figure}(c); Fig.~\ref{outbreak_figure}(c)). Such behaviour is similar to the threshold-switching of optimal isolation strategy observed in \citep{hansen2011optimal}, and also the transition of optimal testing from clinical to mixed strategy noticed in \citep{calabrese2022optimal}.

In the long-term scenario, our results demonstrate that prioritization of transmission-reducing interventions should increase in a monotonic fashion with their effectiveness (Fig.~\ref{long-term_figure}(a)). This is suggestive of the fact that in the case of the availability of highly efficient disease prevention measures, it is always advantageous to allocate maximum resources into it. Interestingly, the allocation of recovery-improving interventions does not always behave in the similar way. { Due to the constant influx of susceptible populations, there is an increased risk of new infections when prevention programs are less effective. Thus, when treatment programs are sufficiently effective, i.e., beyond a threshold, the optimal solution is to divert some resources to prioritization of prevention programs. However, this non-monotonic behaviour disappears if the prevention programs are sufficiently effective (Fig.~\ref{long-term_figure}(b)).}

In the outbreak case, our results demonstrate a much simpler optimal resource allocation strategy. In both the transmission-reducing and recovery-improving cases, increasing the effectiveness of an intervention implies the need to allocate more resources to that intervention measure (Fig.~\ref{outbreak_figure}). { This indicates that allocating more resources towards intervention strategies like symptom monitoring program which was reported to be very effective in preventing diseases like Ebola and SARS ~\citep{peak2017comparing}, might prove to be a more fruitful strategy compared to investing in new healthcare facilities during future outbreaks. On the other hand, when treatment programs are highly effective, since the proportion of susceptibles can only decrease over subsequent times, it is unlikely that the risk of new infections will increase. 
Hence, resources need not be diverted to prioritize prevention programs in such a scenario. 
}

Appropriate allocation of limited resources during the early days of an outbreak is capable of reducing the disease burden. The large change in the relative increase of peak size in the case of more effective control strategies highlights the opportunity to reduce disease burden with the help of timely intervention (Fig.~\ref{delay_figure}(a),(b)). {When the intervention measures are less effective, any delay in implementation does not influence the optimal strategy. However, when the interventions are highly effective, although there is a large increase in peak size for too much delay, there can be a wider range of optimal strategies (Fig.~\ref{delay_figure}(c)).}

{Determining the optimal strategy also requires taking into account the nature of returns from investments into intervention measures. When both the prevention and recovery programs give increasing or constant returns to scale, it is always optimal to invest entire resources into either one of them. On the other hand, decreasing returns from one or both intervention programs increases the likelihood of the optimal strategy being sharing of resources (Fig.~\ref{long-term_9_figure}). This indicates that when the prevention programs concern measures like contact tracing which gives decreasing returns \citep{armbruster2007contact}, a fraction of the resources must also be invested into treatment programs. These findings are comparable to earlier studies on HIV where the allocation strategy "all-or-nothing" in case of increasing and constant, and resource sharing in case of decreasing returns to scale were found to be optimal \citep{brandeau2003resource,alistar2014hiv,brandeau2009optimal}. Additionally, for certain effectiveness of interventions, decreasing returns may help achieve a stable disease-free state or initial decay of outbreak size, which may not be possible otherwise (Fig.~\ref{long-term_9_figure}, Fig.~\ref{outbreak_9_figure} in \ref{Appendix-B}). This is because decreasing returns to scale imply relatively lower transmission and higher recovery rates for the same investment as compared to the other return types. On increasing effectiveness further, there is an increased freedom to choose an optimal strategy to drive the system beyond the $\mathcal{R}_0^c=1$ or $S_0\mathcal{R}_0^{\rm out}=1$ line.}




Summing up, this work provides a comprehensive understanding of how the effectiveness of different control measures determines the optimal strategy for resource allocation. In spite of our minimalistic approach, our findings provide fundamental insight into the resource allocation problem in case of infectious disease spread. { A key limitation of our study is the assumption that increasing investment of resources into intervention programs can alter the transmission and recovery rate instantaneously. However, in reality, that most likely is not the case. Additionally, factors like the number of doctors or success of prevention programs will largely depend on societal conditions which are ignored in this study. Further, while we took into account the different types of production functions to determine the optimal strategies, we did not consider functions which may have decreasing return to scale followed by increasing return to scale or vice versa \citep{brandeau2009optimal,brandeau2003resource}}. Albeit we acknowledge that it is not possible to make recommendations on policy design using a such simple framework, our model does provide a foundational basis on which more specific disease models may be studied in the future. The results highlight the importance of more systematic and involved analyses of the trade-offs between different available strategies by policymakers, especially in developing countries.    

\appendix
\section{Analysis for long-term dynamics}\label{Appendix-A}
$\bullet$~\textbf{{Calculations for $u^*_{\rm long}$ corresponding to asymptotically stable endemic state$\colon$}}\\
  To control for long-term disease persistence, when there is an {asymptotically stable endemic state ($\displaystyle \mathcal{R}_0^c> 1$)}, we differentiate $I^*$ with respect to $u$ and set it equal to zero. Then after simplifying we obtain $u_{\rm l}$ as given in Eq.~(\ref{u}).\\
 At $u = u_{\rm l}$, we have $\displaystyle\frac{d^2 I^*}{d u^2} =\displaystyle \frac{2\mu\beta_e}{\beta_0}\sqrt{\frac{\gamma_0\beta_e\gamma_e}{\beta_0}} > 0$. This shows that $I^*$ is minimum at $u = u_{\rm l}$.\\
 Since the fraction of allocated resources must be constrained between $0$ and $1$, from Eq.~\eqref{u}, we have the inequality
 \begin{equation}
   \displaystyle0<\beta_e(\gamma_0+\gamma_0\gamma_e+\mu)-\sqrt{\beta_0\beta_e\gamma_0\gamma_e}< \beta_e\gamma_0\gamma_e. 
   \label{inequality_long-term}
 \end{equation}
 {For each given $\gamma_e$, from the left part of the inequality \eqref{inequality_long-term}, we get a threshold $\beta_e$ \Big($\displaystyle\beta_e^l=\frac{\beta_0\gamma_0\gamma_e}{(\gamma_0+\gamma_0\gamma_e+\mu)^2}$\Big) such that if $\displaystyle\beta_e\leq \beta_e^l$, we have $u^*_{\rm long}=0$ and if $\displaystyle\beta_e> \beta_e^l$, we have $u^*_{\rm long}>0$ (see Fig.~\ref{long-term_figure}(a)). Note that $\beta_e^l$ is non-monotonic in $\gamma_e$ which we can also visualize in Fig.~\ref{long-term_figure}(a) and Fig.~\ref{long-term_figure}(c). Moreover, for $\beta_e>\beta_e^l$, we get $\displaystyle\frac{\partial u}{\partial \beta_e} =\displaystyle \frac{1}{2}\sqrt{\frac{\beta_0}{\gamma_0\gamma_e{\beta_e}^3}} > 0$.\\
 For each given $\beta_e$, from the right part of the inequality \eqref{inequality_long-term}, we get a threshold $\gamma_e$ \Big($\displaystyle \gamma_e^l=\frac{\beta_e(\gamma_0+\mu)^2}{\beta_0\gamma_0}$\Big) such that if $\displaystyle\gamma_e\leq \gamma_e^l$, we get $u^*_{\rm long}=1$ and if $\displaystyle\gamma_e>\gamma_e^l$, we get $u^*_{\rm long}<1$ (see Fig.~\ref{long-term_figure}(b)). Moreover, for $\displaystyle\gamma_e>\gamma_e^l$, we get $\displaystyle\frac{\partial u}{\partial \gamma_e} =\displaystyle \frac{1}{2}\sqrt{\frac{\beta_0}{\gamma_0\beta_e{\gamma_e}^3}} - \frac{\gamma_0+\mu}{\gamma_0{\gamma_e}^2}$. Now, $\displaystyle\frac{\partial u}{\partial \gamma_e} = 0$ implies $\displaystyle\widetilde{\gamma_e}=\frac{4\beta_e(\gamma_0+\mu)^2}{\beta_0\gamma_0}$ such that for $\displaystyle\widetilde{\gamma_e}>\gamma_e>\gamma_e^l$, we have $\displaystyle\frac{\partial u}{\partial \gamma_e} < 0$ and for $\displaystyle\gamma_e>\widetilde{\gamma_e}>\gamma_e^l$, we have $\displaystyle\frac{\partial u}{\partial \gamma_e} > 0$ (see Fig.~\ref{long-term_figure}(b)).}\\

\begin{figure}
    	\begin{center}
    		\includegraphics[width=0.9\textwidth]{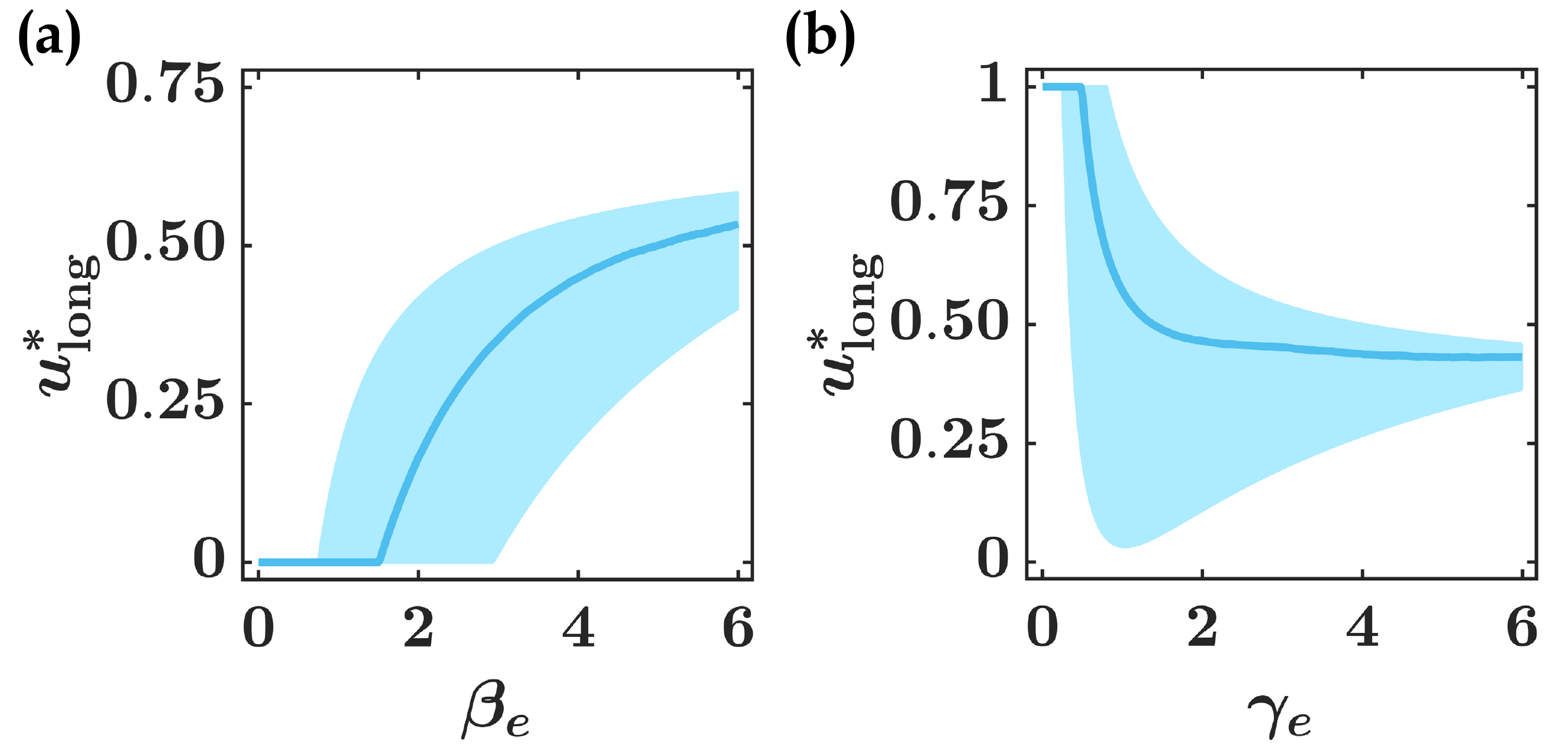}
    	\end{center}
    	\caption{\rm { Sensitivity of (a) $\beta_e$ vs. $u^*_{\rm long}$ (for $\gamma_e=3$), (b) $\gamma_e$ vs. $u^*_{\rm long}$ (for $\beta_e=4$) to the baseline model parameters. The shaded area represents all the obtained values of $u^*_{\rm long}$ when the baseline parameters are varied: $\beta_0\in [0.5, 1.1]$, $\gamma_0\in [0.07, 0.13]$ and $\mu\in[\frac{1}{80}, \frac{1}{60}]$. The solid line represents the median.}}
    	\label{sensitivity_figure}
    \end{figure}

$\bullet$~\textbf{Sensitivity analysis$\colon$}\\
{To explore the sensitivity of our findings to the baseline model parameters ($\beta_0$, $\gamma_0$ and $\mu$), we vary $\beta_0$ from 0.5 to 1.1 $day^{-1}$, $\gamma_0$ from 0.07 to 0.13 $day^{-1}$, and $\mu$ from $\frac{1}{80}$ to $\frac{1}{60}$ $year^{-1}$. Using the Latin Hypercube Sampling scheme, we draw 10,000 samples and obtain 10,000 values of $u^*_{\rm long}$ for a particular $\beta_e$, $\gamma_e$. Then we plot $\beta_e$ vs. $u^*_{\rm long}$ (for $\gamma_e=3$) and $\gamma_e$ vs. $u^*_{\rm long}$ (for $\beta_e=4$) with each of the samples (see Fig.~\ref{sensitivity_figure}).}\\
$\bullet$~\textbf{Non-unique $u^*_{\rm long}$ corresponding to the stable disease-free state$\colon$}\\
 {As the number of infected individuals asymptotically approaches zero whenever $\displaystyle \mathcal{R}_0^c \leq 1$ (Eq.~\eqref{I_last}), solving the inequality $\mathcal{R}_0^c\leq1$ for $u$ will generate the optimal fraction of resources.
Using expression of $\mathcal{R}_0^c$, we have

\begin{equation*}
    \displaystyle \frac{\beta_0}{(1+ \beta_e u)(\gamma_0(1+ \gamma_e (1-u))+\mu)}\leq1.
\end{equation*}}

After simplifying, we get a quadratic inequality of $u$ as\\ $\displaystyle Au^2 + Bu +C \leq0 $, where
$ A = \displaystyle\gamma_0\gamma_e\beta_e$, $B =\displaystyle \gamma_0\gamma_e(1-\beta_e)-\beta_e(\gamma_0+\mu)$, $C=\displaystyle \beta_0-(\gamma_0+\mu)-\gamma_0\gamma_e$.\\
Now, solving the quadratic inequality, we get an interval for $u$ as

 \begin{equation*}
     \displaystyle \big[u_1, u_2\big] =\displaystyle \Bigg[\frac{-B-\sqrt{B^2-4AC}}{2A}, \frac{-B+\sqrt{B^2-4AC}}{2A}\Bigg]
  \end{equation*}
 {Since the fraction of allocated resources must be constrained between $0$ and $1$, we have the optimal fraction of resources corresponding to $\mathcal{R}_0^c\leq1$ lies within
  \begin{equation}
      \displaystyle \big[u_{\rm l_1}, u_{\rm l_2}\big]= [u_1, u_2]\cap[0, 1]
      \label{u_long_interval}
  \end{equation}
 Selecting any $u\in\displaystyle \big[u_{\rm l_1}, u_{\rm l_2}\big]$ would give $\mathcal{R}_0^c\leq1$ (as $\displaystyle\big[u_{\rm l_1}, u_{\rm l_2}\big]\subseteq [u_1, u_2]$), and consequently from Eq.~\eqref{I_last} we have $I^*=0$. Conversely, $I^*=0$ implies $\mathcal{R}_0^c\leq1$, and consequently $u\in\displaystyle \big[u_{\rm l_1}, u_{\rm l_2}\big]$. That means the interior points of the interval \eqref{u_long_interval} are the values of $u^*_{\rm long}$ for which $I^*=0$ with particular $\beta_e$, $\gamma_e$.} That is why $u^*_{\rm long}$ is not unique within \boxed{4} in Fig.~\ref{long-term_figure}(c). 

\begin{figure}
    	\begin{center}
    		\includegraphics[ width=0.9\textwidth]{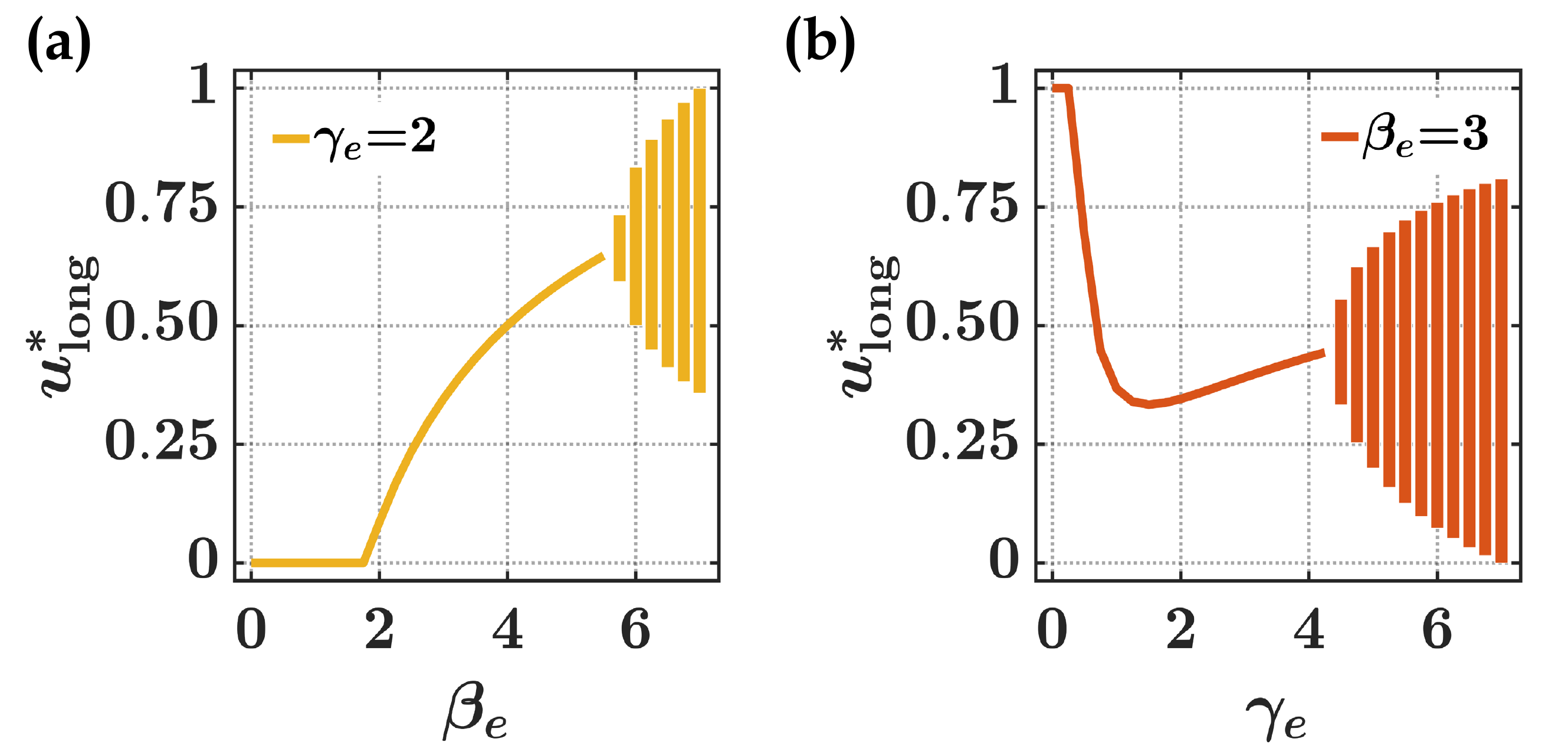}
    	\end{center}
    	\caption{\rm{ The unique and non-unique optimal fraction of resources ($u^*_{\rm long}$) given by Eq.~\eqref{u_long} for the case of long-term disease dynamics. The vertical lines denote the range of possible values for $u^*_{\rm long}$ in the non-unique case. Here unique $u^*_{\rm long}$ corresponds to \boxed{1},\boxed{2} and \boxed{3} while non-unique $u^*_{\rm long}$ corresponds to \boxed{4} in Fig.~\ref{long-term_figure}(c). (a) $\beta_e$ vs. $u^*_{\rm long}$ for $\gamma_e=2$ and (b) $\gamma_e$ vs. $u^*_{\rm long}$ for $\beta_e=3$. Rest of the parameters are taken from Table \ref{parameters_table}. 
     }}
   \label{unique_non_unique_u_long_figure}
    \end{figure}

{Additionally, for specific $\beta_e$ and $\gamma_e$, in Fig.~\ref{unique_non_unique_u_long_figure}, we display unique and non-unique $u^*_{\rm long}$ to control long-term disease persistence associated with both the stable equilibria states.}
\begin{prop}\label{proposition-1}
     {For the resource allocation in long-term disease dynamics scenario, if both $\displaystyle F_{\beta_0}$ and $\displaystyle F_{\gamma_0}$ are with increasing or constant returns to scale or one with increasing and the other constant returns to scale, the optimal fraction of resource $u^*_{\rm long}$ only takes the extreme values (either 0 or 1) within the region of the asymptotically stable endemic state. Moreover, $u^*_{\rm long}$ must be the same in all these cases for a given $\beta_e$, $\gamma_e$.}
\end{prop}
\textbf{Proof:}
    {\noindent In general, during the control of long-term dynamics of disease with any type of returns to scale for $F_{\beta_0}$ and $F_{\gamma_0}$, from Eq.~\eqref{I_last} we have
    
\begin{equation*}
        \displaystyle I^*=\mu \Big(\frac{1}{(F_{\gamma_0}+\mu)}-\frac{1}{F_{\beta_0}}\Big) \quad \big(~\text{as}~ \mathcal{R}_0^c=\frac{F_{\beta_0}}{F_{\gamma_0}+\mu}~\big)
    \end{equation*}
    corresponding to an asymptotically stable endemic state ($\mathcal{R}_0^c>1$). Here both $F_{\beta_0}$, $F_{\gamma_0}$ are differentiable on $(0, 1)$ and $F_{\beta_0}, F_{\gamma_0}\neq0$ on $[0, 1]$. Consequently, $I^*$ is also differentiable on $(0, 1)$.\\
    Now, differentiating $I^*$ with respect to $u$ we have
    \begin{eqnarray}
        \displaystyle I^{*'}&=& \mu \Bigg(\frac{F'_{\beta_0}(F_{\gamma_0}+\mu)^2 - F'_{\gamma_0}(F_{\beta_0})^2}{(F_{\beta_0})^2(F_{\gamma_0}+\mu)^2} \Bigg),
        \label{I^*'}
    \end{eqnarray}
    and
    \begin{equation}
        \displaystyle I^{*''}=\mu \Bigg(\frac{2(F'_{\gamma_0})^2}{(F_{\gamma_0}+\mu)^3}-\frac{2(F'_{\beta_0})^2}{(F_{\beta_0})^3} -\frac{F''_{\gamma_0}}{(F_{\gamma_0}+\mu)^2}+ \frac{F''_{\beta_0}}{(F_{\beta_0})^2} \Bigg).
        \label{I^*''}
    \end{equation}}
    
    $\bullet$ \textbf{Case-1: Both the production functions $\displaystyle F_{\beta_0}$ and $\displaystyle F_{\gamma_0}$ with constant returns to scale}\\
    {As $\displaystyle F_{\beta_0}(u)$ has constant returns to scale, then $\displaystyle F_{\beta_0}(u)$ is linearly decreasing function of $u$ (i.e., $\displaystyle F'_{\beta_0}<0$ and $\displaystyle F''_{\beta_0}=0$). Also, since $\displaystyle F_{\gamma_0}(1-u)$ has constant returns to scale, $\displaystyle F_{\gamma_0}(1-u)$ is linearly increasing function of $(1-u)$, and equivalently, $\displaystyle F_{\gamma_0}$ is linearly decreasing function of $u$ (i.e., $\displaystyle F'_{\gamma_0}<0$ and $\displaystyle F''_{\gamma_0}=0$).\\
   It is to be noted that we denote $I^*$ as $I^*_{\rm CC}$ when both production functions yield constant returns to scale. From Eq.~\eqref{I^*''}, we have
   
   \begin{equation*}
        \displaystyle I^{*''}_{\rm CC}=2\mu \Bigg(\frac{(F'_{\gamma_0})^2}{(F_{\gamma_0}+\mu)^3}-\frac{(F'_{\beta_0})^2}{(F_{\beta_0})^3} \Bigg).
    \end{equation*}
If $I^*_{\rm CC}$ is convex on any subinterval $S$ of $[0, 1]$ (i.e., $I^{*''}_{\rm CC}\geq0$ for all $u\in S\subseteq [0, 1]$), then we have
\begin{eqnarray}
    \nonumber\displaystyle \frac{F'_{\gamma_0}}{F'_{\beta_0}}&\geq& \Bigg(\frac{F_{\gamma_0}+\mu}{F_{\beta_0}} \Bigg)^{3/2}\\
    \implies (\mathcal{R}_0^c)^{3/2}F'_{\gamma_0} &\leq& F'_{\beta_0}, ~\text{for all}~u\in S,
    \label{appendix_inequality-1}
\end{eqnarray}

because $F'_{\beta_0}<0$ and $F'_{\gamma_0}<0$.\\
Now, for all $u\in S\subseteq [0, 1]$,
\begin{eqnarray}
\begin{array}{lll}
    && F'_{\beta_0}(F_{\gamma_0}+\mu)^2 - F'_{\gamma_0}(F_{\beta_0})^2\\\\
    &\geq& (\mathcal{R}_0^c)^{3/2}F'_{\gamma_0}(F_{\gamma_0}+\mu)^2 - F'_{\gamma_0}(F_{\beta_0})^2\\
    && \hspace{3cm}(\text{using Eq.}~\eqref{appendix_inequality-1})\\\\
    &=& (-F'_{\gamma_0})(F_{\gamma_0}+\mu)^2 \Bigg(\frac{(F_{\beta_0})^2}{(F_{\gamma_0}+\mu)^2}-(\mathcal{R}_0^c)^{3/2} \Bigg)\\\\
    &=& (-F'_{\gamma_0})(F_{\gamma_0}+\mu)^2 \Big((\mathcal{R}_0^c)^2 - (\mathcal{R}_0^c)^{3/2} \Big)\\\\
    &>&0 \quad (\text{as}~ F'_{\gamma_0}<0,~ \mathcal{R}_0^c>1)
    \end{array}
    \label{appendix_inequality-2}
\end{eqnarray}

From Eq.~\eqref{I^*'} we have $I^{*'}_{\rm CC}\geq0$ for all $u\in S\subseteq [0, 1]$. So, we get $I^*_{\rm CC}$ as an increasing function of $u$ on $S\subseteq [0, 1]$ whenever $I^*_{\rm CC}$ is convex on $S\subseteq [0, 1]$. Thus if $I^*_{\rm CC}$ is convex on a subinterval of $[0, 1]$, then $I^*_{\rm CC}$ must be increasing on that subinterval. Also, it is worth mentioning that $I^*_{\rm CC}$ may be convex or concave entirely on [0, 1]. In that case, it is obvious to see that $I^*_{\rm CC}$ attains its minimum either at $0$ and $1$. Moreover, there may be situations where $I^*_{\rm CC}$ is first convex on a certain subinterval and then concave on the rest of the subinterval of $[0, 1]$. It is important to point out that if $I^*_{\rm CC}$ is first concave on a certain subinterval of $[0, 1]$, then it cannot be further convex on the remaining subinterval of $[0, 1]$. Otherwise, there must be a cusp point of $I^*_{\rm CC}$ within $(0, 1)$ where $I^*_{\rm CC}$ fails to be differentiable, which would contradict the differentiability of $I^*_{\rm CC}$ on $(0, 1)$.

Here we are trying to minimize $I^*_{\rm CC}$ over the interval $[0, 1]$. If $I^*_{\rm CC}$ is convex entirely on [0, 1], then $I^*_{\rm CC}$ must attain its minimum value at $0$ as $I^*_{\rm CC}$ increases over $[0, 1]$ in that case. If $I^*_{\rm CC}$ is concave entirely on [0, 1] or first convex on a certain subinterval and then concave on the remaining subinterval of [0, 1], then $I^*_{\rm CC}$ must achieve its minimum value either at $0$ (if $I^*_{\rm CC}(u)|_{u=0}<I^*_{\rm CC}(u)|_{u=1}$) or $1$ (if $I^*_{\rm CC}(u)|_{u=0}>I^*_{\rm CC}(u)|_{u=1}$).\\
If possible let $I^*_{\rm CC}$ attain its minimum value at a point $u=c$ other than $0, 1$, so $c\in(0, 1)$.\\
Since $I^*_{\rm CC}$ is minimum at $u=c\in(0, 1)$ and $I^*_{\rm CC}$ is differentiable on $(0, 1)$, then the following must hold in a certain neighborhood of $c$ :

\begin{equation*}
I^{*'}_{\rm CC}(u)=\begin{cases}
          <0 \quad &\text{as} \, u\rightarrow c^- \\
          0 \quad &\text{at} \, u=c \\
          >0 \quad &\text{as} \, u\rightarrow c^+. \\
     \end{cases}
\end{equation*}
This means that there must be a certain neighborhood of $c$ at which $I^{*}_{\rm CC}$ becomes convex and non-monotonic. This contradicts the fact that $I^{*}_{\rm CC}$ must increase if convex. Thus our assumption is incorrect and $I^{*}_{\rm CC}$ cannot attain its minimum value at any point $c\in(0, 1)$. Consequently, $I^{*}_{\rm CC}$ attains minimum value only at 0 or 1. Therefore, $u^*_{\rm long}$ only takes the extreme values (either 0 or 1) within the region of the asymptotically stable endemic state.}\\

\noindent$\bullet$ \textbf{Case-2: Both the production functions $\displaystyle F_{\beta_0}$ and $\displaystyle F_{\gamma_0}$ with increasing returns to scale}\\
{As both $\displaystyle F_{\beta_0}(u)$ and $\displaystyle F_{\gamma_0}(1-u)$ have increasing returns to scale, then $\displaystyle F_{\beta_0}(u)$ is concave (concave down) and decreasing function of $u$ (i.e., $\displaystyle F'_{\beta_0}<0$ and $\displaystyle F''_{\beta_0}\leq0$). Also, $\displaystyle F_{\gamma_0}(1-u)$ is convex (concave up) and increasing function of $(1-u)$, equivalently, $\displaystyle F_{\gamma_0}$ is convex and decreasing function of $u$ (i.e., $\displaystyle F'_{\gamma_0}<0$ and $\displaystyle F''_{\gamma_0}\geq0$).\\
Here we denote $I^*$ as $I^*_{\rm II}$ when both production functions yield increasing returns. From Eq.~\eqref{I^*''}, we have
   \begin{eqnarray*}
\nonumber\displaystyle I^{*''}_{\rm II}&=& \mu \Bigg(\frac{2(F'_{\gamma_0})^2}{(F_{\gamma_0}+\mu)^3}-\frac{2(F'_{\beta_0})^2}{(F_{\beta_0})^3} -\frac{F''_{\gamma_0}}{(F_{\gamma_0}+\mu)^2}+ \frac{F''_{\beta_0}}{(F_{\beta_0})^2} \Bigg)\\
&\leq& 2\mu\Bigg(\frac{(F'_{\gamma_0})^2}{(F_{\gamma_0}+\mu)^3}-\frac{(F'_{\beta_0})^2}{(F_{\beta_0})^3} \Bigg).
   \end{eqnarray*}
With similar arguments as Eq.~\eqref{appendix_inequality-1}, Eq.~\eqref{appendix_inequality-2} in case-1, here we can prove that $I^*_{\rm II}$ is increasing in $u$ whenever $I^*_{\rm II}$ is convex, and consequently, $u^*_{\rm long}$ only takes the extreme values (either 0 or 1) within $\mathcal{R}_0^c>1$.}\\

\noindent$\bullet$ \textbf{Case-3: One of the production functions $\displaystyle F_{\beta_0}$ and $\displaystyle F_{\gamma_0}$ with increasing and the other with constant returns to scale}\\
{First, we consider that $\displaystyle F_{\beta_0}(u)$ and $\displaystyle F_{\gamma_0}(1-u)$ have increasing and constant returns to scale respectively. Then we have $\displaystyle F'_{\beta_0}<0$, $\displaystyle F''_{\beta_0}\leq0$, $\displaystyle F'_{\gamma_0}<0$ and $\displaystyle F''_{\gamma_0}=0$.\\
Here we denote $I^*$ as $I^*_{\rm IC}$ for $\displaystyle F_{\beta_0}$ and $\displaystyle F_{\gamma_0}$ with increasing and constant returns to scale respectively. From Eq.~\eqref{I^*''}, we have
   \begin{eqnarray*}
       \nonumber\displaystyle I^{*''}_{\rm IC}&=&\mu \Bigg(\frac{2(F'_{\gamma_0})^2}{(F_{\gamma_0}+\mu)^3}-\frac{2(F'_{\beta_0})^2}{(F_{\beta_0})^3} + \frac{F''_{\beta_0}}{(F_{\beta_0})^2} \Bigg)\\
       &\leq& 2\mu\Bigg(\frac{(F'_{\gamma_0})^2}{(F_{\gamma_0}+\mu)^3}-\frac{(F'_{\beta_0})^2}{(F_{\beta_0})^3} \Bigg).
   \end{eqnarray*}
   Likewise, we have  \begin{eqnarray*}
       \nonumber\displaystyle I^{*''}_{\rm CI}&=&\mu \Bigg(\frac{2(F'_{\gamma_0})^2}{(F_{\gamma_0}+\mu)^3}-\frac{2(F'_{\beta_0})^2}{(F_{\beta_0})^3} -\frac{F''_{\gamma_0}}{(F_{\gamma_0}+\mu)^2} \Bigg)\\
       &\leq& 2\mu\Bigg(\frac{(F'_{\gamma_0})^2}{(F_{\gamma_0}+\mu)^3}-\frac{(F'_{\beta_0})^2}{(F_{\beta_0})^3} \Bigg).
   \end{eqnarray*}
In this case, also using similar arguments as Eq.~\eqref{appendix_inequality-1}, Eq.~\eqref{appendix_inequality-2} in case-1, it can be shown that $u^*_{\rm long}$ only takes the extreme values either 0 or 1 within $\mathcal{R}_0^c>1$.}\\

{Here we consider three types of returns to scale for $\displaystyle F_{\beta_0}$ (and $\displaystyle F_{\gamma_0}$) which have identical values at extreme point 0 or 1 for certain $\beta_e$, $\gamma_e$ (i.e., three return types for $\displaystyle F_{\beta_0}$ (and $\displaystyle F_{\gamma_0}$) are equivalent to each other whenever u = 0 or u = 1). This is why the value of the objective function, $I^*$ with any type of returns to scale related to $\displaystyle F_{\beta_0}$ and $\displaystyle F_{\gamma_0}$ must be identical at the two extreme points for a given $\beta_e$ and $\gamma_e$. Since $u^*_{\rm long}$ only takes extreme values within $\mathcal{R}_0^c>1$, $u^*_{\rm long}$ must be the same in all these cases described above for a given $\beta_e$, $\gamma_e$. For this reason, regions \boxed{1}, \boxed{3} in the corresponding sub-figures ((a), (b), (d), (e)) in Fig.~\ref{long-term_9_figure} are identical.}

\section{Analysis for outbreak scenario}
\label{Appendix-B}
$\bullet$~\textbf{Derivation of the epidemic peak ($I_{\rm max}$) $\colon$}\\
{From the first two equations of the outbreak model (i.e., model (\ref{model_with_intervention}) with $\mu =0 $) we get 

\begin{eqnarray}
    \displaystyle dI/dt =\displaystyle \Bigg(-1 + \frac{1}{S(t)\mathcal{R}_0^{\rm out}}\Bigg) dS/dt.
    \label{di_dt}
    \end{eqnarray}
    
We know that $dS/dt<0$ at all times in the outbreak scenario for non-zero initial infection sizes ($I_0 = I(0)$). Then from Eq.~\eqref{di_dt}, we observe that $dI/dt$ is initially greater than, equal to, or less than zero when $\displaystyle S_0\mathcal{R}_0^{\rm out}$ is greater than, equal to, or less than one, respectively. Thus the outbreak is initially in a growing, stationary, or decaying state when $\displaystyle S_0\mathcal{R}_0^{\rm out}$ is greater than, equal to, or less than one, respectively. Also, if $dI/dt<0$ at some time $t_1$, then $dI/dt<0$ at any later time $t_2$. This means, if the outbreak size begins to decay at any time, it continues to decay for all subsequent periods.}\\ 
Now, integrating the above Eq.~\eqref{di_dt} and using the initial condition we get

  \begin{equation*}
      I(t) =\displaystyle I_0 + S_0 - S(t) + \frac{1}{\mathcal{R}_0^{\rm out}}\ln{\frac{S(t)}{S_0}.}
  \end{equation*}
  
{Now, if the outbreak is initially in a growing or stationary state ($\displaystyle S_0\mathcal{R}_0^{\rm out}\geq 1$), during the peak of the epidemic (let the peak be reached at $t=t_{\rm peak}$), the rate at which individuals are infected becomes zero (i.e., $dI/dt = 0$ at $t=t_{\rm peak}$). Then from Eq.~\eqref{di_dt}, we get $\displaystyle S(t_{\rm peak}) = \displaystyle\frac{1}{\mathcal{R}_0^{\rm out}}$ (i.e., the fraction of susceptible is equal to the reciprocal of control reproduction number similar as in \citep{keeling2011modeling}).\\ 
Thus from above equation, we obtain the epidemic peak
\begin{eqnarray}
\begin{array}{cc}
  \displaystyle I_{\rm max}=I(t_{\rm peak})=\displaystyle S_0+I_0-\frac{\bigg(\ln{\Big(S_0\mathcal{R}_0^{\rm out}\Big)}+1\bigg)}{\mathcal{R}_0^{\rm out}}.
\end{array}
\end{eqnarray}

From the above expression, we have $I_{\rm max} =I_0$ for the special case $\displaystyle S_0\mathcal{R}_0^{\rm out} = 1$. Also, for $\displaystyle S_0\mathcal{R}_0^{\rm out} < 1$, we must have $I_{\rm max} =I_0$ because $dI/dt<0$ at all times. Finally, the expression for the epidemic peak related to several initial states of the outbreak size is given by Eq.~\eqref{I_max}}.\\
 
$\bullet$~\textbf{Derivation of $u^*_{\rm peak}$ corresponding to initial growth of the outbreak size $\colon$}\\
 {For the outbreak scenarios, when the outbreak is initially in a growing state ($\displaystyle S_0\mathcal{R}_0^{\rm out}> 1$), resources are distributed in such a way that the epidemic peak ($I_{\rm max}$) will be minimum.} Differentiating $I_{\rm max}$ with respect to $u$ and after simplification we have \\
 \begin{eqnarray*}
  \displaystyle \frac{dI_{\rm max}}{du} =\displaystyle \frac{\ln{(S_0\mathcal{R}_0^{\rm out})}}{{\Big(\mathcal{R}_0^{\rm out}\Big)}^2}  \frac{d\mathcal{R}_0^{\rm out}}{du}.\\ 
 \end{eqnarray*}
 Equating it to zero and simplifying we obtain $\displaystyle u_{\rm p} =\displaystyle\frac{1}{2}\Big(1-\frac{1}{\beta_e}+\frac{1}{\gamma_e}\Big)$. At $u=u_{\rm p}$, we have $\displaystyle\frac{d^2 I_{\rm max}}{d u^2} = \displaystyle\frac{2\gamma_0\beta_e\gamma_e}{\beta_0}\ln{(S_0\mathcal{R}_0^{\rm out})} > 0$. This shows that $I_{\rm max}$ is minimum at $u = u_{\rm p}$.\\
 
$\bullet$~\textbf{Effect of $\beta_e, \gamma_e$ on $u^*_{\rm peak}$ during the initial growth of the outbreak state (i.e., within $S_0\mathcal{R}_0^{\rm out}>1$) $\colon$} \\
For extremely lower $\beta_e$, $u^*_{\rm peak}$ tends to 0 and for extremely lower $\gamma_e$, $u^*_{\rm peak}$ tends to 1 respectively. Also, since the fraction of allocated resources must be constrained between $0$ and $1$, from Eq.~\eqref{u_p}, we have the inequality
\begin{equation}
   -\beta_e\gamma_e< \beta_e-\gamma_e< \beta_e\gamma_e.\\
   \label{inequality_outbreak}
 \end{equation}
 {For each given $\gamma_e$, from the left part of the inequality \eqref{inequality_outbreak}, we get a threshold $\beta_e$ \Big($\displaystyle\beta_e^p=\frac{\gamma_e}{1+\gamma_e}$\Big) such that if $\displaystyle\beta_e\leq \beta_e^p$, we have $u^*_{\rm peak}=0$ and if $\displaystyle\beta_e> \beta_e^p$, we have $u^*_{\rm peak}>0$ (see Fig.~\ref{outbreak_figure}(a)). Moreover, for $\beta_e>\beta_e^p$, we get $\displaystyle\frac{\partial u_{\rm p}}{\partial \beta_e} = \displaystyle\frac{1}{2{\beta_e}^2} > 0$.\\
 For each given $\beta_e$, from the right part of the inequality \eqref{inequality_outbreak}, we get a threshold $\gamma_e$ \Big($\displaystyle \gamma_e^p=\frac{\beta_e}{1+\beta_e}$\Big) such that if $\displaystyle\gamma_e\leq \gamma_e^p$, we get $u^*_{\rm peak}=1$ and if $\displaystyle\gamma_e>\gamma_e^p$, we get $u^*_{\rm peak}<1$ (see Fig.~\ref{outbreak_figure}(b)). Moreover, for $\displaystyle\gamma_e>\gamma_e^p$, we get $\displaystyle\frac{\partial u_{\rm p}}{\partial \gamma_e} =\displaystyle -\frac{1}{2{\gamma_e}^2} < 0$.}\\

Now, from the expression for $\displaystyle u_{\rm p}$, we have the following relationship between $u_{\rm peak}^{*}$ and $(1-u_{\rm peak}^{*})$ for all $\beta_e, \gamma_e$ as
\begin{equation*}
\displaystyle u_{\rm peak}^{*}(\beta_e, \gamma_e) = 1- u_{\rm peak}^{*}(\gamma_e, \beta_e).\\
\end{equation*}
\begin{equation*}\centering
  \displaystyle \text{i.e.,}~ u_{\rm peak}^{*}(\beta_e, \gamma_e) +u_{\rm peak}^{*}(\gamma_e, \beta_e)=1.
\end{equation*}
Utilizing the fact $\displaystyle u^*_{\rm peak}=\frac{1}{2}$ for each $\beta_e=\gamma_e$, from above relation we can observe that $u_{\rm peak}^{*}$ partitioned the parameter space of $\beta_e, \gamma_e$ into two parts along the line $\beta_e=\gamma_e$ such that within the parameter space $\displaystyle\{(\beta_e, \gamma_e): \beta_e>\gamma_e\}$, we get $u_{\rm peak}^{*}>\frac{1}{2}$, on the other hand, $u_{\rm peak}^{*}<\frac{1}{2}$ within $\displaystyle\{(\beta_e, \gamma_e): \beta_e<\gamma_e\}$. So we can conclude that $u^*_{\rm peak}$ show symmetric behaviour about the line $\beta_e=\gamma_e$ within the parameter space of $\beta_e, \gamma_e$ during initial growth of the outbreak size.\\

 $\bullet$~{\textbf{Non-unique $u^*_{\rm peak}$ corresponding to the initial decay of the outbreak size$\colon$}\\
 Since the outbreak is initially in a stationary or decaying state (i.e., $I_{\rm max}=I_0$) according to $S_0\mathcal{R}_0^{\rm out}=1$ or $S_0\mathcal{R}_0^{\rm out}<1$ respectively, solving the inequality $S_0\mathcal{R}_0^{\rm out}\leq1$ for $u$ would generate the optimal fraction of resources. Using the expression of $\mathcal{R}_0^{\rm out}$ we have
 
 \begin{eqnarray*}
    \frac{S_0\beta_0}{\gamma_0(1+\beta_eu)(1+\gamma_e(1-u))}\leq1. 
 \end{eqnarray*}
 Simplifying algebraically we get a quadratic inequality of $u$ as $Au^2 + Bu +C \leq0 $.} Here $A= \beta_e\gamma_e$, $B= \gamma_e-\beta_e-\beta_e\gamma_e$ and $C=\displaystyle \frac{S_0\beta_0}{\gamma_0}-\gamma_e-1$. Now, solving the quadratic inequality, we get an interval for $u$ as
 
 {\begin{equation*}
     \displaystyle \big[u_a, u_b\big] =\displaystyle \Bigg[\frac{-B-\sqrt{B^2-4AC}}{2A}, \frac{-B+\sqrt{B^2-4AC}}{2A}\Bigg]
  \end{equation*}
 Since the fraction of allocated resources must be constrained between $0$ and $1$, we have the optimal fraction of resources corresponding to $S_0\mathcal{R}_0^{\rm out}\leq1$ lies within
  \begin{equation}
      \displaystyle \big[u_{\rm p_1}, u_{\rm p_2}\big]= [u_a, u_b]\cap[0, 1] \label{u_peak_interval}
  \end{equation}
  Selecting any $u\in\displaystyle \big[u_{\rm p_1}, u_{\rm p_2}\big]$ would give $S_0\mathcal{R}_0^{\rm out}\leq1$ (as $\displaystyle\big[u_{\rm p_1}, u_{\rm p_2}\big]\subseteq [u_a, u_b]$), and consequently from Eq.~\eqref{I_max} we have $I_{\rm max}=I_0$ (i.e., the outbreak being unable to grow initially). Conversely, $I_{\rm max}=I_0$ implies $S_0\mathcal{R}_0^{\rm out}\leq1$, and consequently $u\in\displaystyle \big[u_{\rm p_1}, u_{\rm p_2}\big]$. That means the interior points of the interval \eqref{u_peak_interval} are the values of $u^*_{\rm peak}$ for which $I_{\rm max}=I_0$ with particular $\beta_e$, $\gamma_e$. That is why $u^*_{\rm peak}$ is not unique within \boxed{4} in Fig.~\ref{outbreak_figure}(c).}

\begin{figure}
    	\begin{center}
    		 \includegraphics[width=0.95\textwidth]{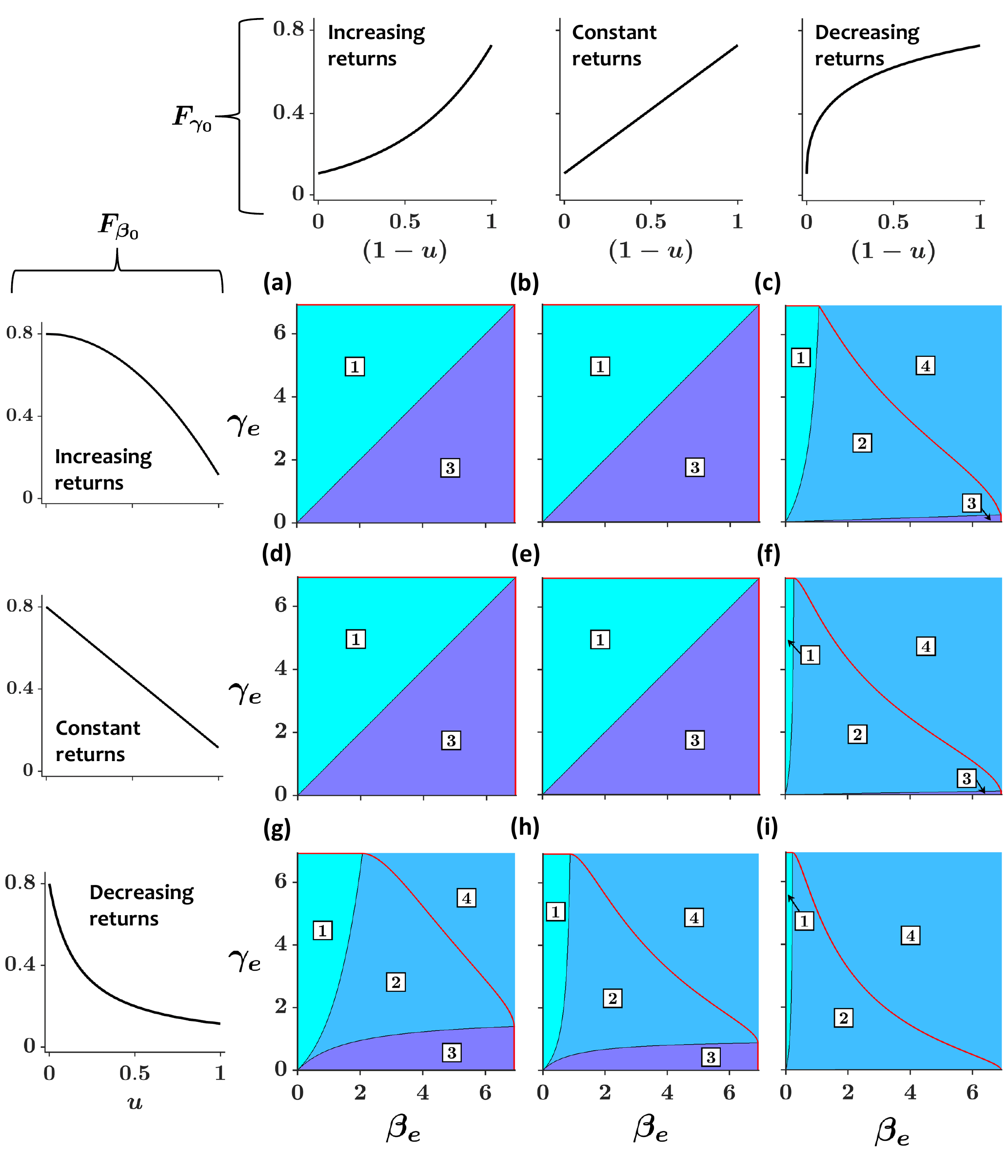}
    	\end{center}
    	\caption{{\rm Comparison of the results considering three types of returns to scale for both $F_{\beta_0}$ and $F_{\gamma_0}$ to explore the dependence of optimal resource allocation on production functions for control of disease in an outbreak scenario. The characteristics of $u^*_{\rm peak}$ within \boxed{1}, \boxed{2}, \boxed{3} and \boxed{4} are the same as described in Fig.~\ref{outbreak_figure}(c). Functional forms for $\displaystyle F_{\beta_0}$ with three types of returns to scale are given as follows: $\displaystyle F_{\beta_0}=\displaystyle\beta_0\Big(1-\frac{\beta_e u^2}{1+\beta_e}\Big)$ (increasing returns to scale), $\displaystyle\beta_0\Big(1-\frac{\beta_e u}{1+\beta_e}\Big)$ (constant returns to scale), and $\displaystyle\frac{\beta_0}{1+\beta_e u}$ (decreasing returns to scale). Also, Functional forms for $\displaystyle F_{\gamma_0}$ with three types of returns to scale are given as follows: $\displaystyle F_{\gamma_0}=\displaystyle\gamma_0\big(1+\gamma_e\big)^{(1-u)}$ (increasing returns to scale), $\displaystyle\gamma_0\big(1+\gamma_e(1-u)\big)$ (constant returns to scale), and $\displaystyle\gamma_0\Bigg(1+\frac{2\gamma_e\sqrt{(1-u)}}{1+\sqrt{(1-u)}}\Bigg)$ (decreasing returns to scale).}
    	}
    	\label{outbreak_9_figure}
    \end{figure}    
    \clearpage
\enlargethispage{20pt}


\section*{Author's contribution}
\textbf{B. M.:} Methodology, Formal analysis, Software, Writing - original draft, Writing - review $\&$ editing. \textbf{S. B.:} Conceptualization, Methodology, Supervision, Writing - review $\&$ editing. \textbf{A. S.:} Conceptualization, Methodology, Supervision, Writing - review $\&$ editing.
\textbf{J. C.:} Supervision, Conceptualization, Writing - review $\&$ editing.


\section*{Acknowledgements}
B. M. is supported by Junior Research Fellowship from University Grants Commission (UGC), India. A. S. and S. B. would like to acknowledge Senior Research Fellowship from CSIR, India for funding them during the initial part of this work. A.S. is also partially funded by the Center of Advanced Systems Understanding (CASUS), which is financed by Germany’s Federal Ministry of Education and Research (BMBF) and by the Saxon Ministry for Science, Culture and Tourism (SMWK) with tax funds on the basis of the budget approved by the Saxon State Parliament. S.B. was supported by the Visiting Scientist fellowship at Indian Statistical Institute, Kolkata during a part of this work. S.B. would also like to acknowledge his present funding under the Marie Skłodowska–Curie grant agreement 101025056 for the project `SpatialSAVE'.




\bibliographystyle{elsarticle-harv} 
\bibliography{main}

\begin{thebibliography}{48}
\expandafter\ifx\csname natexlab\endcsname\relax\def\natexlab#1{#1}\fi
\providecommand{\url}[1]{\texttt{#1}}
\providecommand{\href}[2]{#2}
\providecommand{\path}[1]{#1}
\providecommand{\DOIprefix}{doi:}
\providecommand{\ArXivprefix}{arXiv:}
\providecommand{\URLprefix}{URL: }
\providecommand{\Pubmedprefix}{pmid:}
\providecommand{\doi}[1]{\href{http://dx.doi.org/#1}{\path{#1}}}
\providecommand{\Pubmed}[1]{\href{pmid:#1}{\path{#1}}}
\providecommand{\bibinfo}[2]{#2}
\ifx\xfnm\relax \def\xfnm[#1]{\unskip,\space#1}\fi
\bibitem[{Adhikari et~al.(2020)Adhikari, Meng, Wu, Mao, Ye, Wang, Sun, Sylvia,
  Rozelle, Raat et~al.}]{adhikari2020epidemiology}
\bibinfo{author}{Adhikari, S.P.}, \bibinfo{author}{Meng, S.},
  \bibinfo{author}{Wu, Y.J.}, \bibinfo{author}{Mao, Y.P.}, \bibinfo{author}{Ye,
  R.X.}, \bibinfo{author}{Wang, Q.Z.}, \bibinfo{author}{Sun, C.},
  \bibinfo{author}{Sylvia, S.}, \bibinfo{author}{Rozelle, S.},
  \bibinfo{author}{Raat, H.}, et~al., \bibinfo{year}{2020}.
\newblock \bibinfo{title}{Epidemiology, causes, clinical manifestation and
  diagnosis, prevention and control of coronavirus disease {(COVID-19)} during
  the early outbreak period: a scoping review}.
\newblock \bibinfo{journal}{Infectious Diseases of Poverty}
  \bibinfo{volume}{9}, \bibinfo{pages}{1--12}.
\bibitem[{Alistar et~al.(2014)Alistar, Long, Brandeau and
  Beck}]{alistar2014hiv}
\bibinfo{author}{Alistar, S.S.}, \bibinfo{author}{Long, E.F.},
  \bibinfo{author}{Brandeau, M.L.}, \bibinfo{author}{Beck, E.J.},
  \bibinfo{year}{2014}.
\newblock \bibinfo{title}{{HIV} epidemic control—a model for optimal
  allocation of prevention and treatment resources}.
\newblock \bibinfo{journal}{Health Care Management Science}
  \bibinfo{volume}{17}, \bibinfo{pages}{162--181}.
\bibitem[{Anderson and May(1992)}]{anderson1992infectious}
\bibinfo{author}{Anderson, R.M.}, \bibinfo{author}{May, R.M.},
  \bibinfo{year}{1992}.
\newblock \bibinfo{title}{Infectious diseases of humans: dynamics and control}.
\newblock \bibinfo{publisher}{Oxford university press}.
\bibitem[{Angulo et~al.(2021)Angulo, Casta{\~n}os, Moreno-Morton,
  Velasco-Hern{\'a}ndez and Moreno}]{angulo2021simple}
\bibinfo{author}{Angulo, M.T.}, \bibinfo{author}{Casta{\~n}os, F.},
  \bibinfo{author}{Moreno-Morton, R.}, \bibinfo{author}{Velasco-Hern{\'a}ndez,
  J.X.}, \bibinfo{author}{Moreno, J.A.}, \bibinfo{year}{2021}.
\newblock \bibinfo{title}{A simple criterion to design optimal
  non-pharmaceutical interventions for mitigating epidemic outbreaks}.
\newblock \bibinfo{journal}{Journal of the Royal Society Interface}
  \bibinfo{volume}{18}, \bibinfo{pages}{20200803}.
\bibitem[{Armbruster and Brandeau(2007)}]{armbruster2007contact}
\bibinfo{author}{Armbruster, B.}, \bibinfo{author}{Brandeau, M.L.},
  \bibinfo{year}{2007}.
\newblock \bibinfo{title}{Contact tracing to control infectious disease: when
  enough is enough.}
\newblock \bibinfo{journal}{Health care management science}
  \bibinfo{volume}{10}, \bibinfo{pages}{341--355}.
\bibitem[{Beigel et~al.(2020)Beigel, Tomashek, Dodd, Mehta, Zingman, Kalil,
  Hohmann, Chu, Luetkemeyer, Kline et~al.}]{beigel2020remdesivir}
\bibinfo{author}{Beigel, J.H.}, \bibinfo{author}{Tomashek, K.M.},
  \bibinfo{author}{Dodd, L.E.}, \bibinfo{author}{Mehta, A.K.},
  \bibinfo{author}{Zingman, B.S.}, \bibinfo{author}{Kalil, A.C.},
  \bibinfo{author}{Hohmann, E.}, \bibinfo{author}{Chu, H.Y.},
  \bibinfo{author}{Luetkemeyer, A.}, \bibinfo{author}{Kline, S.}, et~al.,
  \bibinfo{year}{2020}.
\newblock \bibinfo{title}{Remdesivir for the treatment of covid-19}.
\newblock \bibinfo{journal}{New England Journal of Medicine}
  \bibinfo{volume}{383}, \bibinfo{pages}{1813--1826}.
\bibitem[{Benson et~al.(2009)Benson, Brooks, Holmes, Kaplan, Masur and
  Pau}]{benson2009guidelines}
\bibinfo{author}{Benson, C.A.}, \bibinfo{author}{Brooks, J.T.},
  \bibinfo{author}{Holmes, K.K.}, \bibinfo{author}{Kaplan, J.E.},
  \bibinfo{author}{Masur, H.}, \bibinfo{author}{Pau, A.}, \bibinfo{year}{2009}.
\newblock \bibinfo{title}{Guidelines for prevention and treatment opportunistic
  infections in {HIV}-infected adults and adolescents; recommendations from
  cdc, the national institutes of health, and the {HIV} medicine
  association/infectious diseases society of america} .
\bibitem[{Bolzoni et~al.(2019)Bolzoni, Bonacini, Della~Marca and
  Groppi}]{bolzoni2019optimal}
\bibinfo{author}{Bolzoni, L.}, \bibinfo{author}{Bonacini, E.},
  \bibinfo{author}{Della~Marca, R.}, \bibinfo{author}{Groppi, M.},
  \bibinfo{year}{2019}.
\newblock \bibinfo{title}{Optimal control of epidemic size and duration with
  limited resources}.
\newblock \bibinfo{journal}{Mathematical Biosciences} \bibinfo{volume}{315},
  \bibinfo{pages}{108232}.
\bibitem[{Brandeau and Zaric(2009)}]{brandeau2009optimal}
\bibinfo{author}{Brandeau, M.L.}, \bibinfo{author}{Zaric, G.S.},
  \bibinfo{year}{2009}.
\newblock \bibinfo{title}{Optimal investment in hiv prevention programs: more
  is not always better}.
\newblock \bibinfo{journal}{Health care management science}
  \bibinfo{volume}{12}, \bibinfo{pages}{27--37}.
\bibitem[{Brandeau et~al.(2005)Brandeau, Zaric and
  De~Angelis}]{brandeau2005improved}
\bibinfo{author}{Brandeau, M.L.}, \bibinfo{author}{Zaric, G.S.},
  \bibinfo{author}{De~Angelis, V.}, \bibinfo{year}{2005}.
\newblock \bibinfo{title}{Improved allocation of hiv prevention resources:
  using information about prevention program production functions}.
\newblock \bibinfo{journal}{Health Care Management Science}
  \bibinfo{volume}{8}, \bibinfo{pages}{19--28}.
\bibitem[{Brandeau et~al.(2003)Brandeau, Zaric and
  Richter}]{brandeau2003resource}
\bibinfo{author}{Brandeau, M.L.}, \bibinfo{author}{Zaric, G.S.},
  \bibinfo{author}{Richter, A.}, \bibinfo{year}{2003}.
\newblock \bibinfo{title}{Resource allocation for control of infectious
  diseases in multiple independent populations: beyond cost-effectiveness
  analysis}.
\newblock \bibinfo{journal}{Journal of health economics} \bibinfo{volume}{22},
  \bibinfo{pages}{575--598}.
\bibitem[{Breda et~al.(2012)Breda, Diekmann, De~Graaf, Pugliese and
  Vermiglio}]{breda2012formulation}
\bibinfo{author}{Breda, D.}, \bibinfo{author}{Diekmann, O.},
  \bibinfo{author}{De~Graaf, W.}, \bibinfo{author}{Pugliese, A.},
  \bibinfo{author}{Vermiglio, R.}, \bibinfo{year}{2012}.
\newblock \bibinfo{title}{On the formulation of epidemic models (an appraisal
  of {Kermack} and {McKendrick})}.
\newblock \bibinfo{journal}{Journal of Biological Dynamics}
  \bibinfo{volume}{6}, \bibinfo{pages}{103--117}.
\bibitem[{Calabrese and Demers(2022)}]{calabrese2022optimal}
\bibinfo{author}{Calabrese, J.M.}, \bibinfo{author}{Demers, J.},
  \bibinfo{year}{2022}.
\newblock \bibinfo{title}{How optimal allocation of limited testing capacity
  changes epidemic dynamics}.
\newblock \bibinfo{journal}{Journal of Theoretical Biology}
  \bibinfo{volume}{538}, \bibinfo{pages}{111017}.
\bibitem[{Colizza et~al.(2007)Colizza, Barrat, Barth{\'e}lemy and
  Vespignani}]{colizza2007predictability}
\bibinfo{author}{Colizza, V.}, \bibinfo{author}{Barrat, A.},
  \bibinfo{author}{Barth{\'e}lemy, M.}, \bibinfo{author}{Vespignani, A.},
  \bibinfo{year}{2007}.
\newblock \bibinfo{title}{Predictability and epidemic pathways in global
  outbreaks of infectious diseases: the {SARS} case study}.
\newblock \bibinfo{journal}{BMC Medicine} \bibinfo{volume}{5},
  \bibinfo{pages}{1--13}.
\bibitem[{Deka and Bhattacharyya(2019)}]{deka2019game}
\bibinfo{author}{Deka, A.}, \bibinfo{author}{Bhattacharyya, S.},
  \bibinfo{year}{2019}.
\newblock \bibinfo{title}{Game dynamic model of optimal budget allocation under
  individual vaccination choice}.
\newblock \bibinfo{journal}{Journal of Theoretical Biology}
  \bibinfo{volume}{470}, \bibinfo{pages}{108--118}.
\bibitem[{Emanuel et~al.(2020)Emanuel, Persad, Upshur, Thome, Parker, Glickman,
  Zhang, Boyle, Smith and Phillips}]{emanuel2020fair}
\bibinfo{author}{Emanuel, E.J.}, \bibinfo{author}{Persad, G.},
  \bibinfo{author}{Upshur, R.}, \bibinfo{author}{Thome, B.},
  \bibinfo{author}{Parker, M.}, \bibinfo{author}{Glickman, A.},
  \bibinfo{author}{Zhang, C.}, \bibinfo{author}{Boyle, C.},
  \bibinfo{author}{Smith, M.}, \bibinfo{author}{Phillips, J.P.},
  \bibinfo{year}{2020}.
\newblock \bibinfo{title}{Fair allocation of scarce medical resources in the
  time of {COVID-19}}.
\newblock \bibinfo{journal}{New England Journal of Medicine}
  \bibinfo{volume}{382}, \bibinfo{pages}{2049--2055}.
\bibitem[{Farmer(2001)}]{farmer2001major}
\bibinfo{author}{Farmer, P.}, \bibinfo{year}{2001}.
\newblock \bibinfo{title}{The major infectious diseases in the world—to treat
  or not to treat?}
\newblock \bibinfo{journal}{New England Journal of Medicine}
  \bibinfo{volume}{345}, \bibinfo{pages}{208--210}.
\bibitem[{Fraser et~al.(2009)Fraser, Donnelly, Cauchemez, Hanage, Van~Kerkhove,
  Hollingsworth, Griffin, Baggaley, Jenkins, Lyons et~al.}]{fraser2009pandemic}
\bibinfo{author}{Fraser, C.}, \bibinfo{author}{Donnelly, C.A.},
  \bibinfo{author}{Cauchemez, S.}, \bibinfo{author}{Hanage, W.P.},
  \bibinfo{author}{Van~Kerkhove, M.D.}, \bibinfo{author}{Hollingsworth, T.D.},
  \bibinfo{author}{Griffin, J.}, \bibinfo{author}{Baggaley, R.F.},
  \bibinfo{author}{Jenkins, H.E.}, \bibinfo{author}{Lyons, E.J.}, et~al.,
  \bibinfo{year}{2009}.
\newblock \bibinfo{title}{Pandemic potential of a strain of influenza {A}
  {(H1N1)}: early findings}.
\newblock \bibinfo{journal}{Science} \bibinfo{volume}{324},
  \bibinfo{pages}{1557--1561}.
\bibitem[{Ghosh et~al.(2021)Ghosh, Senapati, Chattopadhyay, Hens and
  Ghosh}]{ghosh2021optimal}
\bibinfo{author}{Ghosh, S.}, \bibinfo{author}{Senapati, A.},
  \bibinfo{author}{Chattopadhyay, J.}, \bibinfo{author}{Hens, C.},
  \bibinfo{author}{Ghosh, D.}, \bibinfo{year}{2021}.
\newblock \bibinfo{title}{Optimal test-kit-based intervention strategy of
  epidemic spreading in heterogeneous complex networks}.
\newblock \bibinfo{journal}{Chaos: An Interdisciplinary Journal of Nonlinear
  Science} \bibinfo{volume}{31}, \bibinfo{pages}{071101}.
\bibitem[{Gostin(2014)}]{gostin2014ebola}
\bibinfo{author}{Gostin, L.O.}, \bibinfo{year}{2014}.
\newblock \bibinfo{title}{Ebola: towards an international health systems fund}.
\newblock \bibinfo{journal}{The Lancet} \bibinfo{volume}{384},
  \bibinfo{pages}{e49--e51}.
\bibitem[{Gostin and Friedman(2015)}]{gostin2015retrospective}
\bibinfo{author}{Gostin, L.O.}, \bibinfo{author}{Friedman, E.A.},
  \bibinfo{year}{2015}.
\newblock \bibinfo{title}{A retrospective and prospective analysis of the west
  african ebola virus disease epidemic: robust national health systems at the
  foundation and an empowered who at the apex}.
\newblock \bibinfo{journal}{The Lancet} \bibinfo{volume}{385},
  \bibinfo{pages}{1902--1909}.
\bibitem[{Group(2020)}]{recovery2020effect}
\bibinfo{author}{Group, R.C.}, \bibinfo{year}{2020}.
\newblock \bibinfo{title}{Effect of hydroxychloroquine in hospitalized patients
  with covid-19}.
\newblock \bibinfo{journal}{New England Journal of Medicine}
  \bibinfo{volume}{383}, \bibinfo{pages}{2030--2040}.
\bibitem[{Gubler(2002)}]{gubler2002epidemic}
\bibinfo{author}{Gubler, D.J.}, \bibinfo{year}{2002}.
\newblock \bibinfo{title}{Epidemic dengue/dengue hemorrhagic fever as a public
  health, social and economic problem in the 21st century}.
\newblock \bibinfo{journal}{Trends in microbiology} \bibinfo{volume}{10},
  \bibinfo{pages}{100--103}.
\bibitem[{Hansen and Day(2011)}]{hansen2011optimal}
\bibinfo{author}{Hansen, E.}, \bibinfo{author}{Day, T.}, \bibinfo{year}{2011}.
\newblock \bibinfo{title}{Optimal control of epidemics with limited resources}.
\newblock \bibinfo{journal}{Journal of Mathematical Biology}
  \bibinfo{volume}{62}, \bibinfo{pages}{423--451}.
\bibitem[{Hopman et~al.(2020)Hopman, Allegranzi and
  Mehtar}]{hopman2020managing}
\bibinfo{author}{Hopman, J.}, \bibinfo{author}{Allegranzi, B.},
  \bibinfo{author}{Mehtar, S.}, \bibinfo{year}{2020}.
\newblock \bibinfo{title}{Managing {COVID-19} in low-and middle-income
  countries}.
\newblock \bibinfo{journal}{Jama} \bibinfo{volume}{323},
  \bibinfo{pages}{1549--1550}.
\bibitem[{Hufnagel et~al.(2004)Hufnagel, Brockmann and
  Geisel}]{hufnagel2004forecast}
\bibinfo{author}{Hufnagel, L.}, \bibinfo{author}{Brockmann, D.},
  \bibinfo{author}{Geisel, T.}, \bibinfo{year}{2004}.
\newblock \bibinfo{title}{Forecast and control of epidemics in a globalized
  world}.
\newblock \bibinfo{journal}{Proceedings of the National Academy of Sciences}
  \bibinfo{volume}{101}, \bibinfo{pages}{15124--15129}.
\bibitem[{Kaplan(1995)}]{kaplan1995economic}
\bibinfo{author}{Kaplan, E.H.}, \bibinfo{year}{1995}.
\newblock \bibinfo{title}{Economic analysis of needle exchange}.
\newblock \bibinfo{journal}{AIDs} \bibinfo{volume}{9},
  \bibinfo{pages}{1113--1120}.
\bibitem[{Keeling and Rohani(2011)}]{keeling2011modeling}
\bibinfo{author}{Keeling, M.J.}, \bibinfo{author}{Rohani, P.},
  \bibinfo{year}{2011}.
\newblock \bibinfo{title}{Modeling infectious diseases in humans and animals}.
\newblock \bibinfo{publisher}{Princeton university press}.
\bibitem[{Legrand et~al.(2007)Legrand, Grais, Boelle, Valleron and
  Flahault}]{legrand2007understanding}
\bibinfo{author}{Legrand, J.}, \bibinfo{author}{Grais, R.F.},
  \bibinfo{author}{Boelle, P.Y.}, \bibinfo{author}{Valleron, A.J.},
  \bibinfo{author}{Flahault, A.}, \bibinfo{year}{2007}.
\newblock \bibinfo{title}{Understanding the dynamics of {E}bola epidemics}.
\newblock \bibinfo{journal}{Epidemiology \& Infection} \bibinfo{volume}{135},
  \bibinfo{pages}{610--621}.
\bibitem[{Lima et~al.(2008)Lima, Johnston, Hogg, Levy, Harrigan, Anema and
  Montaner}]{lima2008expanded}
\bibinfo{author}{Lima, V.D.}, \bibinfo{author}{Johnston, K.},
  \bibinfo{author}{Hogg, R.S.}, \bibinfo{author}{Levy, A.R.},
  \bibinfo{author}{Harrigan, P.R.}, \bibinfo{author}{Anema, A.},
  \bibinfo{author}{Montaner, J.S.}, \bibinfo{year}{2008}.
\newblock \bibinfo{title}{Expanded access to highly active antiretroviral
  therapy: a potentially powerful strategy to curb the growth of the hiv
  epidemic}.
\newblock \bibinfo{journal}{The Journal of infectious diseases}
  \bibinfo{volume}{198}, \bibinfo{pages}{59--67}.
\bibitem[{Mahase(2020)}]{mahase2020china}
\bibinfo{author}{Mahase, E.}, \bibinfo{year}{2020}.
\newblock \bibinfo{title}{China coronavirus: {WHO} declares international
  emergency as death toll exceeds 200}.
\newblock \bibinfo{journal}{BMJ: British Medical Journal (Online)}
  \bibinfo{volume}{368}.
\bibitem[{Martcheva(2015)}]{martcheva2015introduction}
\bibinfo{author}{Martcheva, M.}, \bibinfo{year}{2015}.
\newblock \bibinfo{title}{An introduction to mathematical epidemiology}.
  volume~\bibinfo{volume}{61}.
\newblock \bibinfo{publisher}{Springer}.
\bibitem[{Medlock and Galvani(2009)}]{medlock2009optimizing}
\bibinfo{author}{Medlock, J.}, \bibinfo{author}{Galvani, A.P.},
  \bibinfo{year}{2009}.
\newblock \bibinfo{title}{Optimizing influenza vaccine distribution}.
\newblock \bibinfo{journal}{Science} \bibinfo{volume}{325},
  \bibinfo{pages}{1705--1708}.
\bibitem[{Mosadeghrad(2014)}]{mosadeghrad2014factors}
\bibinfo{author}{Mosadeghrad, A.M.}, \bibinfo{year}{2014}.
\newblock \bibinfo{title}{Factors influencing healthcare service quality}.
\newblock \bibinfo{journal}{International Journal of Health Policy and
  Management} \bibinfo{volume}{3}, \bibinfo{pages}{77}.
\bibitem[{Ngonghala et~al.(2020)Ngonghala, Iboi, Eikenberry, Scotch, MacIntyre,
  Bonds and Gumel}]{ngonghala2020mathematical}
\bibinfo{author}{Ngonghala, C.N.}, \bibinfo{author}{Iboi, E.},
  \bibinfo{author}{Eikenberry, S.}, \bibinfo{author}{Scotch, M.},
  \bibinfo{author}{MacIntyre, C.R.}, \bibinfo{author}{Bonds, M.H.},
  \bibinfo{author}{Gumel, A.B.}, \bibinfo{year}{2020}.
\newblock \bibinfo{title}{Mathematical assessment of the impact of
  non-pharmaceutical interventions on curtailing the 2019 novel coronavirus}.
\newblock \bibinfo{journal}{Mathematical biosciences} \bibinfo{volume}{325},
  \bibinfo{pages}{108364}.
\bibitem[{Oshitani et~al.(2008)Oshitani, Kamigaki and
  Suzuki}]{oshitani2008major}
\bibinfo{author}{Oshitani, H.}, \bibinfo{author}{Kamigaki, T.},
  \bibinfo{author}{Suzuki, A.}, \bibinfo{year}{2008}.
\newblock \bibinfo{title}{Major issues and challenges of influenza pandemic
  preparedness in developing countries}.
\newblock \bibinfo{journal}{Emerging Infectious Diseases} \bibinfo{volume}{14},
  \bibinfo{pages}{875}.
\bibitem[{Peak et~al.(2017)Peak, Childs, Grad and Buckee}]{peak2017comparing}
\bibinfo{author}{Peak, C.M.}, \bibinfo{author}{Childs, L.M.},
  \bibinfo{author}{Grad, Y.H.}, \bibinfo{author}{Buckee, C.O.},
  \bibinfo{year}{2017}.
\newblock \bibinfo{title}{Comparing nonpharmaceutical interventions for
  containing emerging epidemics}.
\newblock \bibinfo{journal}{Proceedings of the National Academy of Sciences}
  \bibinfo{volume}{114}, \bibinfo{pages}{4023--4028}.
\bibitem[{Sacchetto et~al.(2020)Sacchetto, Raviolo, Beltrando and
  Tommasoni}]{sacchetto2020covid}
\bibinfo{author}{Sacchetto, D.}, \bibinfo{author}{Raviolo, M.},
  \bibinfo{author}{Beltrando, C.}, \bibinfo{author}{Tommasoni, N.},
  \bibinfo{year}{2020}.
\newblock \bibinfo{title}{{COVID-19} surge capacity solutions: Our experience
  of converting a concert hall into a temporary hospital for mild and moderate
  {COVID-19} patients}.
\newblock \bibinfo{journal}{Disaster Medicine and Public Health Preparedness} ,
  \bibinfo{pages}{1--4}.
\bibitem[{Senapati et~al.(2019)Senapati, Sardar, Ganguly, Ganguly,
  Chattopadhyay and Chattopadhyay}]{senapati2019impact}
\bibinfo{author}{Senapati, A.}, \bibinfo{author}{Sardar, T.},
  \bibinfo{author}{Ganguly, K.S.}, \bibinfo{author}{Ganguly, K.S.},
  \bibinfo{author}{Chattopadhyay, A.K.}, \bibinfo{author}{Chattopadhyay, J.},
  \bibinfo{year}{2019}.
\newblock \bibinfo{title}{Impact of adult mosquito control on dengue prevalence
  in a multi-patch setting: a case study in {Kolkata} (2014--2015)}.
\newblock \bibinfo{journal}{Journal of Theoretical Biology}
  \bibinfo{volume}{478}, \bibinfo{pages}{139--152}.
\bibitem[{Siegel et~al.(2007)Siegel, Rhinehart, Jackson and
  Chiarello}]{siegel20072007}
\bibinfo{author}{Siegel, J.D.}, \bibinfo{author}{Rhinehart, E.},
  \bibinfo{author}{Jackson, M.}, \bibinfo{author}{Chiarello, L.},
  \bibinfo{year}{2007}.
\newblock \bibinfo{title}{2007 guideline for isolation precautions: preventing
  transmission of infectious agents in health care settings}.
\newblock \bibinfo{journal}{American journal of infection control}
  \bibinfo{volume}{35}, \bibinfo{pages}{S65--S164}.
\bibitem[{WHO(2005)}]{world2005checklist}
\bibinfo{author}{WHO}, \bibinfo{year}{2005}.
\newblock \bibinfo{title}{Who checklist for influenza pandemic preparedness
  planning}.
\bibitem[{WHO(2020)}]{world2020coronavirus}
\bibinfo{author}{WHO}, \bibinfo{year}{2020}.
\newblock \bibinfo{title}{Coronavirus disease 2019 {(COVID-19)}: situation
  report, 73}.
\bibitem[{Worby and Chang(2020)}]{worby2020face}
\bibinfo{author}{Worby, C.J.}, \bibinfo{author}{Chang, H.H.},
  \bibinfo{year}{2020}.
\newblock \bibinfo{title}{Face mask use in the general population and optimal
  resource allocation during the {COVID-19} pandemic}.
\newblock \bibinfo{journal}{Nature communications} \bibinfo{volume}{11},
  \bibinfo{pages}{1--9}.
\bibitem[{Zaric et~al.(2000a)Zaric, Barnett and Brandeau}]{zaric2000hiv}
\bibinfo{author}{Zaric, G.S.}, \bibinfo{author}{Barnett, P.G.},
  \bibinfo{author}{Brandeau, M.L.}, \bibinfo{year}{2000}a.
\newblock \bibinfo{title}{Hiv transmission and the cost-effectiveness of
  methadone maintenance.}
\newblock \bibinfo{journal}{American journal of public health}
  \bibinfo{volume}{90}, \bibinfo{pages}{1100}.
\bibitem[{Zaric and Brandeau(2001)}]{zaric2001optimal}
\bibinfo{author}{Zaric, G.S.}, \bibinfo{author}{Brandeau, M.L.},
  \bibinfo{year}{2001}.
\newblock \bibinfo{title}{Optimal investment in a portfolio of hiv prevention
  programs}.
\newblock \bibinfo{journal}{Medical Decision Making} \bibinfo{volume}{21},
  \bibinfo{pages}{391--408}.
\bibitem[{Zaric et~al.(2000b)Zaric, Brandeau and Barnett}]{zaric2000methadone}
\bibinfo{author}{Zaric, G.S.}, \bibinfo{author}{Brandeau, M.L.},
  \bibinfo{author}{Barnett, P.G.}, \bibinfo{year}{2000}b.
\newblock \bibinfo{title}{Methadone maintenance and hiv prevention: a
  cost-effectiveness analysis}.
\newblock \bibinfo{journal}{Management Science} \bibinfo{volume}{46},
  \bibinfo{pages}{1013--1031}.
\bibitem[{Zhang et~al.(2020a)Zhang, Litvinova, Liang, Wang, Wang, Zhao, Wu,
  Merler, Viboud, Vespignani et~al.}]{zhang2020changes}
\bibinfo{author}{Zhang, J.}, \bibinfo{author}{Litvinova, M.},
  \bibinfo{author}{Liang, Y.}, \bibinfo{author}{Wang, Y.},
  \bibinfo{author}{Wang, W.}, \bibinfo{author}{Zhao, S.}, \bibinfo{author}{Wu,
  Q.}, \bibinfo{author}{Merler, S.}, \bibinfo{author}{Viboud, C.},
  \bibinfo{author}{Vespignani, A.}, et~al., \bibinfo{year}{2020}a.
\newblock \bibinfo{title}{Changes in contact patterns shape the dynamics of the
  {COVID-19} outbreak in {C}hina}.
\newblock \bibinfo{journal}{Science} \bibinfo{volume}{368},
  \bibinfo{pages}{1481--1486}.
\bibitem[{Zhang et~al.(2020b)Zhang, Li, Zhang, Wang and
  Molina}]{zhang2020identifying}
\bibinfo{author}{Zhang, R.}, \bibinfo{author}{Li, Y.}, \bibinfo{author}{Zhang,
  A.L.}, \bibinfo{author}{Wang, Y.}, \bibinfo{author}{Molina, M.J.},
  \bibinfo{year}{2020}b.
\newblock \bibinfo{title}{Identifying airborne transmission as the dominant
  route for the spread of covid-19}.
\newblock \bibinfo{journal}{Proceedings of the National Academy of Sciences}
  \bibinfo{volume}{117}, \bibinfo{pages}{14857--14863}.

\end{thebibliography}
\end{document}